\documentclass[12pt]{iopart}
\usepackage{graphicx}
\usepackage{cite}
\usepackage{dcolumn}
\usepackage{bm}
\usepackage{color}
\usepackage{iopams} 
\usepackage{amscd} 

\begin{document}

\title[Viscoelastic subdiffusion in disordered systems]{Finite-range viscoelastic subdiffusion 
in disordered systems with inclusion of inertial effects}
\author{Igor Goychuk and Thorsten P\"oschel}
\address{Institute for Multiscale Simulation, Friedrich-Alexander University 
of Erlangen-N\"urnberg, Cauerstr. 3,  91058 Erlangen, Germany}
\ead{igor.goychuk@fau.de}


\date{\today}

\begin{abstract}    
This work justifies further paradigmatic importance of the model of
viscoelastic subdiffusion in random environments for the observed subdiffusion in cellular
biological systems. Recently, we showed [PCCP, \textbf{20}, 24140 (2018)] that this model
displays several remarkable features, which makes it an attractive paradigm to explain the
physical nature of subdiffusion occurring in biological cells. In particular, it combines
viscoelasticity with distinct non-ergodic features. We extend this basic model to make it
suitable for physical phenomena such as subdiffusion of lipids in disordered biological
membranes upon including the inertial effects.  For lipids, the inertial effects occur in the
range of picoseconds, and a power-law decaying viscoelastic memory extends over the range of
several nanoseconds.  Thus, in the absence of disorder, diffusion would become normal on a time
scale beyond this memory range. However, both experimentally and in some molecular-dynamical
simulations, the time range of lipid subdiffusion extends far beyond the viscoelastic memory
range. We study three 1d models of correlated quenched Gaussian disorder to explain the puzzle:
singular short-range (exponentially correlated), smooth short-range (Gaussian-correlated), and
smooth long-range (power-law correlated) disorder. For a moderate disorder strength,  transient
viscoelastic subdiffusion changes into the subdiffusion caused by the randomness of the
environment. It is characterized by a time-dependent power-law exponent of subdiffusion
$\alpha(t)$, which can show nonmonotonous behavior, in agreement with some recent
molecular-dynamical simulations. Moreover, the spatial distribution of test particles in this
disorder-dominated regime is shown to be a non-Gaussian, exponential power distribution with
index $\chi=1.45-2.3$, which also correlates well with  molecular-dynamical findings and
experiments. Furthermore, this subdiffusion is nonergodic with single-trajectory averages
showing a broad scatter, in agreement with experimental observations for viscoelastic
subdiffusion of various particles in living cells.   
\end{abstract}


%
\vspace{2pc}
%
%

\maketitle


\section{Introduction}

Nano- and submicron test particles frequently manifest a subdiffusive behavior on experimentally relevant time scales, which range
from picoseconds to hours, in complex polymer and colloidal fluids \cite{MasonPRL,AmblardPRL,WongPRL,
Waigh,Santamaria,WeissPRE13}, optically created random potentials \cite{EversReview,Hanes12,HanesPRE},  
cytosol of biological cells \cite{Saxton97,WeissBJ04,Tolic,BanksBJ05,Golding,
Guigas,SzymanskiPRL,WeigelPNAS,HoflingReview,Jeon11,Luby,PanPRL,Harrison,Parry,Robert,Bertzeva12,Regner13,Manzo15,Lampo17},  
biological membranes \cite{Schwille99,KnellerJCP11,Sezgin12,JeonPRL12, JeonPRX,Metzler16}, and DNA tracks
\cite{WangPRL,GoychukPRL14, Kong16, Kong17,Liu17}. 
Subdiffusion means that the ensemble-averaged variance of the particles position scales
sublinearly in time, $\langle \delta x^2(t)\rangle\propto t^{\alpha}$, $0<\alpha<1$. We shall consider a
one-dimensional case for simplicity. Typically, $\alpha(t)$ is time-dependent and reaches asymptotically
unity, i.e., subdiffusion is a transient phenomenon. However, the pertinent transient time scales can be crucial for
physical problems considered. Also conformational diffusion in biological macromolecules such as proteins is
often anomalously slow \cite{Yang03,GoychukPRE04,Kneller04,
KouPRL,MinPRL,Calladrini10,Calligari11,Calligari15,GoychukPRE15,HuNatPhys16}.  Different theories were
proposed to explain the origin of such subdiffusion based on: (i) continuous-time random walks (CTRWs)
\cite{Shlesinger74,Scher75,Hughes,AvrahamHavlin,Metzler01} with infinite
\cite{Metzler01,Yang03,MetzlerPCCP,Barkai,GoychukPRE04,HePRL08,Jeon11} and finite
\cite{GoychukPRL03,GoychukPRE12,GoychukCTP14} mean residence times (MRTs) in traps and associated fractional
Fokker-Planck equation descriptions  \cite{Metzler99,Metzler01,GoychukPRE06Rapid, GoychukPRE12}; (ii)
viscoelasticity of the environment \cite{MasonPRL,AmblardPRL,Waigh,Guigas, GoychukPRE07Rapid, Santamaria,
GoychukPRE09,Calladrini10,Calligari11,GoychukACP12,FNL12}, or intrinsic viscoelasticity of macromolecules
\cite{Kneller04,KouPRL, MinPRL, GoychukPRL07,GoychukPRL19Reply,GoychukPRE19} based on a generalized Langevin
equation (GLE) description \cite{Kubo66,Zwanzig73, KuboBook,ZwanzigBook}, including fractional Langevin
equations (FLEs) \cite{MainardiFLE,LutzFLE,GoychukPRL07,GoychukPRL19Reply}, and fractional Brownian motion
(FBM) \cite{Kolmogorov,KolmogorovTrans, Mandelbrot68}; (iii) memoryless diffusion in random Gaussian
potentials \cite{Romero98,KhouryPRL,LindenbergFNL,SimonPRE13,HanesPRE,GoychukPRL14,Goychuk2017}; (iv) normal diffusion
coefficient fluctuating in time due to a random environment \cite{Massignan14,Manzo15}; (v) diffusion on
fractal structures \cite{AvrahamHavlin}, and this is not a complete list \cite{HoflingReview,MetzlerPCCP}.
The physical mechanism at play is often controversial, and the experiments are open for concurrent
interpretations. For example, the experiments \cite{Golding} on RNA messengers subdiffusion in E.coli
cells were first interpreted in the framework of a CTRW description with infinite MRTs \cite{HePRL08}.
However, later on, it has been shown that the basic underlying mechanism is viscoelastic and is rather
related to FBM \cite{MagdziarzPRL, Lampo17}. A mixture of different mechanisms can be at play
\cite{TabeiPNAS, WeigelPNAS, PCCP18}. Moreover, many biological cells are expected to make reversible
transitions between a solid-like anabiotic state with frozen metabolic activity, where the CTRW-like
mechanism can be more appropriate to a functionally active fluid-like state with viscoelasticity playing a
leading role \cite{Parry}.

The viscoelastic nature of subdiffusive processes in the cytosol of biological cells in its functional
liquid-like state received a solid experimental support
\cite{Caspi,Golding,Guigas,Waigh,WeberPRL,WeissBJ04,TabeiPNAS,WeigelPNAS,Lampo17,Robert,Bruno11,Harrison}.
It was, however, also challenged \cite{HePRL08,Jeon11,LubelskiPRL08,Barkai} due to a broad distribution
of diffusion coefficients derived from single-trajectory averages because free viscoelastic subdiffusion
is ergodic \cite{Deng09,GoychukPRE09,GoychukACP12,MetzlerPCCP}. Hence, diffusion coefficient derived from
single-trajectories should not be broadly distributed unless the test particles are broadly distributed
in sizes, if diffusion is ergodic. Moreover, the single-trajectory and ensemble averages should coincide.
This is, however, frequently not the case. One needs to explain a combination of viscoelastic
subdiffusion with non-ergodic features like a large scattering of single-trajectory averages.  For living
cells, it seems obvious that viscoelastic subdiffusion, as a basic pertinent mechanism, should be
combined with trapping caused by elements of the cytoskeleton and molecular crowding. This general idea
was until now expressed within three different modeling approaches: (i) by a combination of FBM with CTRW
via a subordination, i.e., introducing a randomized time in FBM \cite{TabeiPNAS}; (ii) by considering
FBM  living on a fractal structure \cite{WeigelPNAS}, and (iii) by considering viscoelastic GLE
subdiffusion in random potentials \cite{PCCP18} modeling a quenched disorder. The third approach appeals
as more physical and fundamental, allowing for various modifications and huge variations, including
natural generalizations beyond thermal equilibrium \cite{GoychukACP12}. 

Indeed, locally memoryless diffusion in disordered media modeled by a random potential or
random force-field provides the foundation of several anomalous diffusion theories \cite{Bouchaud1990, Hughes}. For
example, the model of CTRW with power-law distributed residence time and divergent MRT in traps comes
naturally from the model of exponential energy $U(x)$ disorder with energy fluctuations characterized by a root-mean-square (RMS) amplitude $\sigma=\langle \delta U^2(x)\rangle^{1/2}$, yielding
$\alpha=k_BT/\sigma<1$ for $\sigma>k_BT$ \cite{Shlesinger74,Scher75,Metzler01}. Furthermore, uncorrelated
random Gaussian force $f(x)$ model, $\langle f(x)f(x')\rangle\propto \delta(x-x')$ yields a Brownian motion
of the potential $U(x)=-\int^x f(x')dx'$ in space, with growing variance, $\langle \delta U^2(x)\rangle
\propto x$. It provides a continuous space generalization of the original lattice Sinai model \cite{Sinai82} as a
Langevin-equation-based subdiffusion in such a fluctuating potential. It yields $\langle \delta x^2(t)
\rangle\propto |\ln(t/t_0)|^a $ with $a=4$ \cite{Bouchaud1990,BouchaudAnnPhys90,Doussal99}, i.e., a logarithmically slow subdiffusion. 

The model of Gaussian energy disorder naturally emerges due to the central limit theorem
\cite{Bouchaud1990} in many physical systems ranging from organic photoconductors
\cite{BasslerReview,DunlapPRL96} and supercooled liquids \cite{BasslerPRL87,HecksherNatPhys} to diffusion
of regulatory proteins on DNA tracks
\cite{GerlandPNAS,SlutskyPRE,LassigReview,BenichouPRL09,Sheinman2012,GoychukPRL14,Kong16,Kong17}. Here,
Sinai model should be contrasted with the model of uncorrelated or short-range correlated random
stationary Gaussian potentials with a finite energy RMS fluctuation, $\sigma$, and normalized
autocorrelation function $g(x)=\langle U(x'+x) U(x')\rangle/\sigma^2$, which seems to be more appropriate
for such systems. It was, however, long time thought to lead just to normal diffusion with a renormalized
diffusion coefficient $D=D_{\sigma=0}\exp[-(\sigma/k_BT)^2]$
\cite{DeGennes75,BasslerPRL87,BasslerReview,HanggiRevModPhys}. This theoretical result is known as de
Gennes-B\"assler-Zwanzig renormalization of diffusion and rate processes by disorder
\cite{HanggiRevModPhys}. However, numerical results on the overdamped Langevin diffusion
\cite{Romero98,KhouryPRL,LindenbergFNL,SimonPRE13,GoychukPRL14,Goychuk2017} and Monte Carlo jump diffusion
\cite{HanesJPCM,HanesPRE} revealed a typical subdiffusion even for a short-range correlated disorder, like
$g(x)=\exp(-|x|/\lambda)$, on the spatial scale $L_{\rm erg}=2\lambda
(k_BT)^2\exp[(\sigma/k_BT)^2]/\sigma^2\gg \lambda$ \cite{GoychukPRL14,Goychuk2017} strongly exceeding the
correlation length $\lambda$, when $\sigma$ exceeds $k_BT$ by several times. Also experiments with
colloidal particles in optically created random Gaussian potentials confirm the existence of such
transient subdiffusion \cite{Hanes12,HanesPRE}.  This subdiffusion occurs paradoxically much faster in
absolute terms than the normal diffusion expected from the de Gennes-B\"assler-Zwanzig renormalization
\cite{GoychukPRL14,Goychuk2017}. Its non-ergodic origin is by now well understood
\cite{GoychukPRL14,Goychuk2017}. Remarkably, de Gennes-B\"assler-Zwanzig renormalization is mathematically
valid asymptotically for any stationary ergodic Gaussian potential with decaying spatial correlations,
$g(x)\to 0$, $x\to\infty$ \cite{GoychukPRL14}. However, it becomes practically irrelevant and even
misleading for $\sigma\gg k_BT$. Moreover, strikingly enough, transient subdiffusion emerges also for a
potential, which is uncorrelated on the sites of a lattice with a discretization step or lattice period
$\Delta x$ and continuous in between \cite{Goychuk2017}. It can also last very long with $\lambda \to
\Delta x$ in its spatial range estimate \cite{Goychuk2017}. 

Another recent surprise was provided by the emergence of
a Sinai-like logarithmic  diffusion with $b\sim 1-4$ in such stationary Gaussian potentials for $\sigma >
5\; k_BT$ \cite{Goychuk2017}. All these remarkable features were explained within a simple scaling theory
\cite{Goychuk2017} inspired by one developed for the continuous-space Sinai diffusion by Bouchaud \textit{at
al.} \cite{Bouchaud1990,BouchaudAnnPhys90}. It explains also the origin of a very long
transient regime of standard, power-law-scaling ensemble subdiffusion with $\alpha\approx 2k_BT/\sigma_{\rm
eff}$ and $\sigma_{\rm eff}\approx (1.24-1.52)\sigma$ depending on the model of $g(x)$ \cite{Goychuk2017}.
This dependence looks remarkably similar to the case of exponential disorder, and the single-trajectory
averages are also broadly scattered. Generally, $\alpha(t)$ is time-dependent, with an extended (for $\sigma> 3k_BT$) intermediate
regime of $\alpha(t)\approx 2k_BT/\sigma_{\rm eff}=const$, which turns into a very slow logarithmic
increase, $\alpha(t)\sim 2(k_BT/\sigma)^2 \ln(t/t_0)$, while approaching the asymptotic regime of normal
diffusion \cite{Goychuk2017}. For these reasons,
we find frequently subdiffusion in random Gaussian potentials in the case of a sufficiently strong disorder, at odds with the outdated dogma of simple renormalization \cite{HanggiRevModPhys}. In this
respect, it is pertinent to note that for proteins nonspecifically bound to DNA, disorder strength can reach
$\sigma\sim  (4-5)\;k_BT\sim (0.1-0.125)\;{\rm eV}$ at ambient room $T$ 
\cite{GerlandPNAS,LassigReview,Sheinman2012}, i.e., it is not small at all, contrary to what is often
stipulated \cite{SlutskyPRE}. To think in terms of normal diffusion with a renormalized diffusion
coefficient in such a situation is very misleading \cite{GoychukPRL14,Goychuk2017}. Indeed, for a typical
$D_{\sigma=0}=3\;\mu{\rm m^2/s}$ \cite{ElfScience07}, $D_{\sigma}\approx 4\times 10^{-5}\;{\rm nm^2/s}$ for
$\sigma=5\;k_BT$, which means that diffusion over 2 nm would take over one day \cite{Goychuk2017}.
Subdiffusional search is orders of magnitude faster \cite{GoychukPRL14,Kong16,Kong17} than this
renormalization predicts.  In a combination with a local energy bias caused by the disorder correlations
\cite{Goychuk2017}, this paradoxically fast subdiffusion provides a way to resolve long-standing speed-stability paradox
\cite{Sheinman2012}.  Such a local energy bias caused by correlated disorder can indeed direct  search to a target binding site and  greatly accelerate it
\cite{BenichouPRL09,Goychuk2017}.

The first systematic study of overdamped viscoelastic subdiffusion in random Gaussian potentials was done in
Ref. \cite{PCCP18} for two fundamental models of correlations: the already mentioned exponentially decaying
correlations (Ornstein-Uhlenbeck process in space) and power-law decaying correlations with infinite
correlation range. An earlier study was also done in \cite{DuanEPJB} for a peculiar model of $g(x)$ and a
weakly corrugated parabolic potential. One of the significant results of \cite{PCCP18} is that a small
disorder, $\sigma<2\;k_BT$, makes an almost negligible impact on viscoelastic subdiffusion on the ensemble
level, i.e., viscoelasticity wins over disorder. However, the disorder can introduce a substantial scatter
of single-trajectory averages already for $\sigma\sim 1-2\;k_BT$. Another significant result is that for a
stronger $\sigma >2\;k_BT$ disorder such a diffusion starts to look on the ensemble level as diffusion
caused by Gaussian disorder (in the absence if viscoelastic memory effects), i.e., the disorder seems to win
over viscoelasticity. It can easily be mistaken for diffusion caused by CTRW with divergent MRT \cite{PCCP18}. However,
residence times are described by a generalized lognormal distribution, and an appropriate scaling time of
subdiffusion is expressed through the fractional friction coefficient \cite{PCCP18}.  

In this paper, we include inertial effects in viscoelastic motion \cite{GoychukPRE09, GoychukACP12},
which are pertinent, e.g., for subdiffusion in lipid systems \cite{KnellerJCP11, JeonPRL12, Metzler16}.
Next, we are looking for a physical explanation of unusual non-Gaussian, exponential power distribution
$P(x,t)\propto \exp\left (-|x/x_{\chi}(t)|^{\chi(t)}\right )$ of  spreading the test particles positions
revealed in viscoelastic subdiffusion in cytosol of living cells \cite{Lampo17} ($\chi$ was fixed to the
value $\chi=1$ of the Laplace distribution in this paper while fitting the experimental data)    and in
biological membranes with a generally time-dependent power-law exponent $\chi(t)$ $\chi(t)\sim 1.4-2.2$
\cite{JeonPRX,Metzler16} and a time-dependent width $x_{\chi}(t)$. We believe that the medium's disorder
causes it.  Indeed, Langevin's memoryless diffusion in Gaussian potentials yields such a distribution
with $\chi(t)\sim 1.4-2$; see Fig. 2 of Supplementary Material in \cite{GoychukPRL14}. Also experimental
distributions  of colloidal particle displacements in optically created random Gaussian potentials  look
similar, cf. Fig. 11 in \cite{Hanes12}.  For viscoelastic subdiffusion in random Gaussian potentials, it
was, however, not studied until now. Next, viscoelastic subdiffusion in pure homogeneous lipid bilayers
has a finite time range. For example, for the cases studied in Ref. \cite{JeonPRL12} subdiffusion regime
with $\alpha\approx 0.6$ lasted until about 10 ns and then crossed over into normal diffusion. However,
in lipid membranes with cholesterol and disordered lipid gel phases, crossover occurs not to normal
diffusion but rather to another subdiffusion regime with $\alpha\approx 0.8$ that lasts until 100 ns
\cite{JeonPRL12, Metzler16}. It is natural to suppose this regime is due to the medium's disorder, rather
than viscoelasticity, or presents a combination of both.

Moreover, a non-monotonic behavior of $\alpha(t)$ was found in such complex lipid systems crowded besides
with proteins \cite{Java13, Metzler16}. Namely, after reaching a maximum $\alpha(t)$ can drop for a while
and then increase again.  In this paper, we conceive and investigate a pertinent basic model capable to
qualitatively explain these puzzling features as a combination of transient viscoelastic subdiffusion and
subdiffusion caused by the disorder.

\section{Model and Theory}

The model is based on subdiffusion governed by a Kubo-Zwanzig GLE
\cite{Bogolyubov,Ford65,Kubo66,Zwanzig73}
\begin{equation}\label{GLEA}
m\ddot x+\int_{0}^{t}\eta(t-t')\dot x(t')dt'=-dU(x)/dx+\xi(t),
\end{equation}
in a zero-mean random stationary Gaussian potential $U(x)$ characterized by a stationary ACF $g(x)$ and
RMS $\sigma$. It yields a quenched random force,  $f(x)=-dU(x)/dx$. In Eq. (\ref{GLEA}), $m$ is the mass of a particle, $\eta(t)$ is a frictional memory kernel
and $\xi(t)$ is a zero-mean thermal Gaussian noise force, which is completely characterized by its
autocorrelation function (ACF) $\langle \xi(t')\xi(t)\rangle$. The friction and noise are related by the
fluctuation-dissipation relation, $\langle \xi(t')\xi(t)\rangle=k_BT\eta(|t-t'|)$, reflecting thermal
fluctuation-dissipation theorem (FDT) \cite{Kubo66}. The memory kernel is assumed to have a power law
decay,  $\eta(t)=\eta_\alpha \exp(-\nu_h t)/[t^\alpha\Gamma(1-\alpha)]$,  with an exponential memory
cutoff $\nu_h$. Here, $\eta_\alpha$ is the fractional  friction coefficient and $\tau_h=1/\nu_h$ is a
crossover time to normal diffusion,  in the absence of quenched random force, $f(x)=0$. 
Then, the natural time unit is the velocity relaxation time scale
$\tau_v=(m/\eta_\alpha)^{1/(2-\alpha)}$  entering the stationary velocity autocorrelation (VACF), which
in the limit $\tau_h\to\infty$  reads \cite{WeissBook,LutzFLE,GoychukPRE09,GoychukACP12} 
\begin{eqnarray}
  K_v(t)=\langle v(t)v(0)\rangle_{\rm st}=v_T^2 E_{2-\alpha}[-(t/\tau_v)^{2-\alpha}]\;.
  \label{Kv}
 \end{eqnarray}
 It provides a very good approximation within our model for $t\ll \tau_h$ and $\tau_h\gg\tau_v$, which is
assumed. Here, $v_T=\sqrt{k_BT/m}$ is thermal velocity, and $E_a(z)=E_{a,b=1}(z)$ is the Mittag-Leffler function, $E_{a,b}(z)=\sum_{n=0}^{\infty}z^n/\Gamma(an+b)$. The law of diffusion is given by the twice-integrated VACF, 
  \begin{eqnarray}\label{diff}
 \langle \delta x^2(t)\rangle=2v_T^2t^2 E_{2-\alpha,3}[-(t/\tau_v)^{2-\alpha}]\;,
  \end{eqnarray}
for initial Maxwell distribution of velocities and $t\ll \tau_h$. Initially, diffusion is ballistic,
$\langle \delta x^2(t)\rangle\approx v_T^2t^2$, $t\ll \tau_v$. Then, it turns into  subdiffusion,
$\langle \delta x^2(t)\rangle\approx 2D_\alpha t^\alpha/\Gamma(1+\alpha)$, for $t\gg \tau_v$, where
$D_\alpha$ is the fractional diffusion coefficient obeying a generalized Einstein relation,
$D_\alpha=k_BT/\eta_\alpha$, which also expresses FDT. Such a course-grained for $t\gg \tau_v$ process is
nothing else FBM and this FBM description holds for $\tau_v\ll t\ll \tau_h$. For $t\gg \tau_h$, the
normal diffusion regime gradually emerges. In this work, we scale time in units of $\tau_v$, velocity in
$v_T$, length in $x_0=v_T\tau_v$, and energy in $k_BT$.

\begin{figure}[h]
\vspace{0.8cm}
	\centering
	\includegraphics[height=5.4cm]{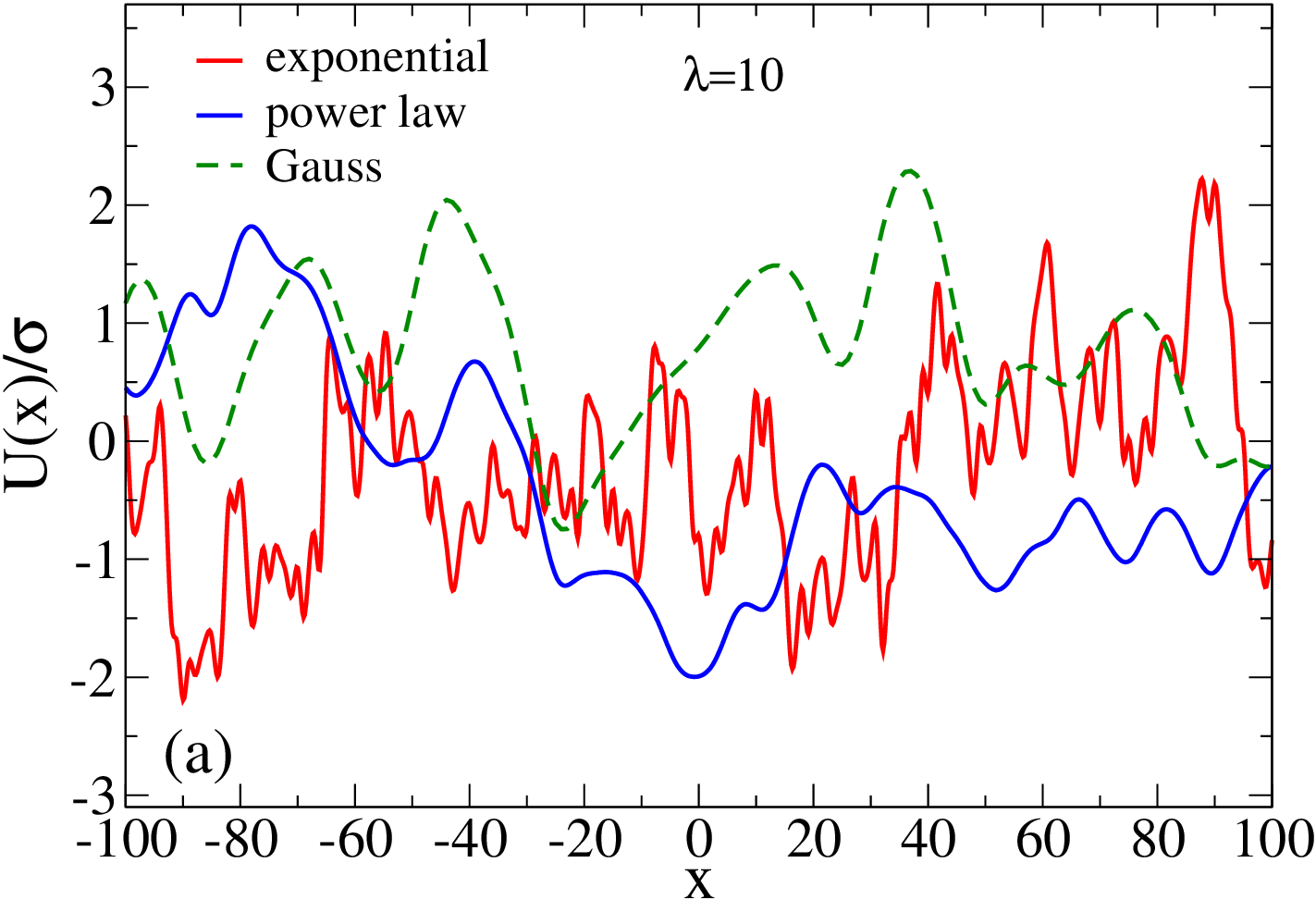}\hfill
\includegraphics[height=5.4cm]{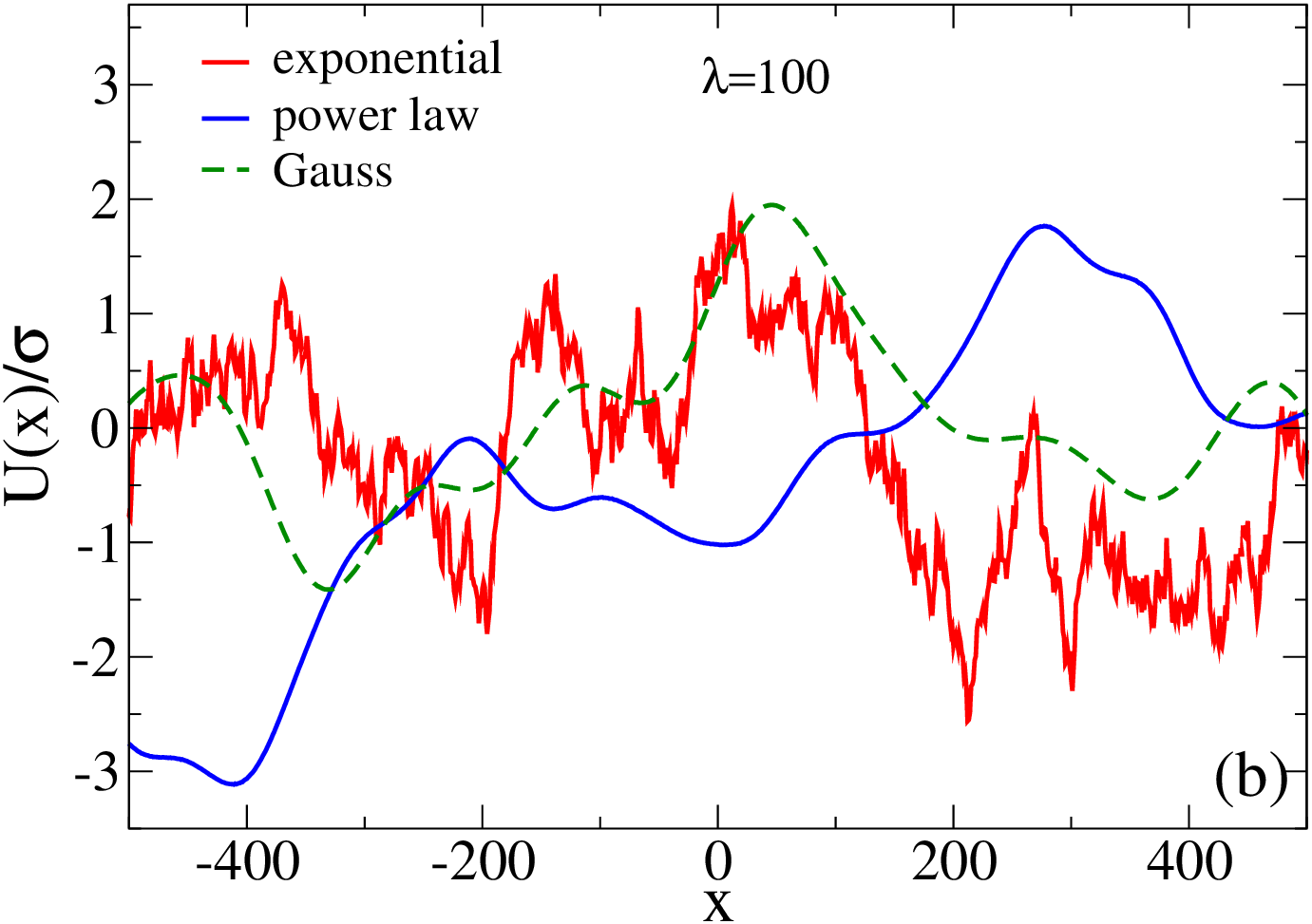}	
	\caption{Realizations of random potentials for exponential,  power-law with $\gamma=1$, and Gauss-decaying correlations. The lattice grid size is $\Delta x=1$. (a) $\lambda=10$; (b) $\lambda=100$. $L=2^{19}\Delta x\approx 5.24\times 10^5$.
	}
	\label{Fig1}       
\end{figure}

The potential is considered on a lattice with spatial grid size $\Delta x$.  It presents randomly
generated energy values on the lattice sites, which are continuously connected by parabolic splines, so
that that the quenched random force $f(x)$ is piece-wise linear in space \cite{SimonFNL}. A spectral
algorithm for the generation of such a random potential is described in \cite{SimonFNL}. It was
successfully used in many papers and requires to specify some large spurious periodization length $L$ and
$g(x)$. We consider three models of correlation decay: (1) $g(x)=\exp(-|x|/\lambda)$ (short-range
singular disorder),  (2) $g(x)=1/[1+x^2/\lambda^2]^{\gamma/2}$ with $\gamma=1$ and $g(x)\sim
(\lambda/|x|)^\gamma\;{\rm at}\;x\gg \lambda$  (long-range non-singular disorder with diverging
correlation length), and (3) $g(x)=\exp(-x^2/\lambda^2)$ (short-range non-singular disorder) for two
values  $\lambda=10$ and $\lambda=100$. The model of power-law correlated disorder with $\gamma=1$
emerges due to a charge-dipole interaction, e.g., in the context of charge diffusion in  molecularly
doped polymers \cite{DunlapPRL96}, where typically $\sigma\sim 4$  in units of room $k_BT_r\sim 0.025$
eV. The electrostatic nature of this model of correlated disorder qualifies it as an important physical
model. The models of exponentially correlated disorder (Ornstein-Uhlenbeck process in space) and Gaussian
correlations are basic models of generic interest. Exponential correlations lead to a singular model,
while a Gaussian decay provides a non-singular model of short-range correlations. Typical realizations of
corresponding random potentials are shown in Fig. \ref{Fig1}.  Notice a very rugged character of
potential fluctuations in the case of exponential correlations and a strong local bias on a typical scale
of $\lambda$ in the cases of exponential and power-law correlations. In this respect, the character of a
smooth disorder is not changed upon making $\Delta x$ ever smaller, provided $\Delta x\ll \lambda$: there
are typically about one or two local potential minima per $\lambda$ length. The  Ornstein-Uhlenbeck process in
space (exponential correlations) is very different: there are huge many local minima and maxima within
its correlation length and their number increases upon diminishing $\Delta x$. Indeed,  this process can
only be defined on a lattice with some finite $\Delta x$ given its singular character because for this
process the force RMS $\langle f^2(x)\rangle^{1/2}=\sigma\sqrt{2/(\Delta x\lambda)}\to\infty $ diverge 
with $\Delta x\to 0$ \cite{GoychukPRL14,Goychuk2017,PCCP18}. Power-law-decaying and Gauss-decaying
correlations yield smooth $U(x)$ \cite{Goychuk2017, PCCP18}. In these cases, $\langle
f^2(x)\rangle^{1/2}=\sqrt{2}\sigma/\lambda$, and  $\langle
f^2(x)\rangle^{1/2}=\sqrt{\gamma}\sigma/\lambda$, respectively, when $\Delta x\to 0$.

Since the analytical treatment of diffusion and transport processes of strongly non-Markovian GLE
dynamics in random correlated potentials is not feasible, we investigated it numerically. 

\subsection{Numerical approach}

The numerical approach is based on approximating the memory kernel by a sum of exponentials and a 
hyper-dimensional Markovian embedding of the low-dimensional GLE dynamics. It is very well developed both
in application to sub-diffusive \cite{GoychukPRE09,GoychukACP12,PCCP18} and super-diffusive GLE
\cite{SiegleEPL11,GoychukPRL19,GoychukPRE20}, including FLE dynamics. This numerically accurate approach
is detailed in \ref{appendA}. 

\subsection{Ensemble vs. trajectory averages, ergodicity and its breaking}

Notice that ergodicity of random processes  can be understood in various senses \cite{Papoulis}, with
ergodicity in the mean value serving as the basic central notion \cite{Papoulis}. The ergodicity of a
diffusion process $x(t)$  is considered in the current literature as ergodicity of $x(t)$ in its squared
increments \cite{MetzlerPCCP}. Namely, it is considered as a mean-value ergodicity of the stochastic
process defined by the squared position increments,  $[\delta x_i(t|t')]^2$, of $x(t)$ within some time
interval $t$, where $\delta x_i(t|t')=x_i(t'+t)-x_i(t')$. Here, $t'$ is the running time of this process
and $t$ is a forward time-increment of the $i$-th trajectory counted from this running time.  The 
time-average of these squared position increments over $t'$ is defined as $\overline{\delta
x_i^2(t)}^{{\cal T}_w}=\frac{1}{{\cal T}_w-t} \int_0^{{\cal T}_w-t}[\delta x_i(t|t')]^2dt'$, i.e., as a
running trajectory average within a time window $\Delta{\cal T}_w  ={\cal T}_w-t\gg t$. It depends on $t$
and ${\cal T}_w$. In 1d, the single-trajectory diffusion coefficient, $D_{i,\rm sgl}$, is defined by
scaling $\overline{\delta x_i^2(t)}^{{\cal T}_w}\sim 2\bar D_{i,\rm sgl}({\cal T}_w) t^{\alpha}$ with
growing $t$. Often, the definition involves an additional gamma-function factor $\Gamma(1+\alpha)$, $\bar
D_{i,\rm sgl}({\cal T}_w)=D_{i,\rm sgl}({\cal T}_w)/\Gamma(1+\alpha)$, which we will, however, not use
for the single-trajectory averages. The empiric ensemble averaging presents a more common concept,
$\langle \delta x^2(t)\rangle_M=(1/M)\sum_{i=1}^M[\delta x_i(t|t')]^2 $, and the ensemble diffusion
coefficient, $D_{M,\alpha}$,  is defined by $ \langle \delta x^2(t)\rangle_M\sim 2\bar D_{M,\alpha}
t^{\alpha}$. $M$ will be mostly omitted in the following while assuming that is sufficiently large.
Moreover, $t'$ is set to zero in the corresponding ensemble averages, while sampling initial particles
positions randomly in space. Notice that only for the stationary in wide sense increments this average
does not depend on $t'$. Diffusion is considered ergodic if both averages exist and coincide,
$\overline{\delta x_i^2(t)}^{\infty}=\langle \delta x^2(t)\rangle_\infty$, in the limit $\Delta {\cal
T}_w\to\infty, M\to\infty$. From a mathematical point of view, these averages can be identical only for
processes with stationary increments. Otherwise, they cannot be expected to coincide \cite{Papoulis,
Yaglom}. Nevertheless, from a physical point of view both averages always make sense and should be
considered different and mutually complementary characteristics of diffusion processes.  Evidently, they
are not expected to ever coincide in disordered media with quenched disorder, where the position
increment for any particular trajectory is never stationary. 

 By Khinchin theorem \cite{Papoulis, Yaglom}, a sufficient condition for ergodicity of a stationary
 random process in the mean value is the decay of its ensemble stationary autocorrelation function (ACF)
 to zero (in the limit $M\to\infty$, which is equivalent to the ensemble-averaging in a proper
 mathematical sense using the probability density function). Slutsky's theorem poses an even weaker
 condition: it is necessary and sufficient that the time-average of the stationary ensemble ACF decays to
 zero in the limit of infinite averaging interval \cite{Papoulis, Yaglom}. For the discussed ergodicity
 of diffusion processes, a more complex sufficient condition necessarily emerges, and it can be
 formulated in terms of four-points correlation functions of increments \cite{Papoulis}. However, for
 Gaussian processes, it reduces to decay of the (infinite) ensemble autocorrelation function of
 increments to zero \cite{Papoulis, Yaglom}. It is the case of FBM (a singular non-differentiable
 stochastic process with stationary increments) and its inertial generalization \cite{Deng09,
 GoychukPRE09} based on GLE, including FLE, with a well defined (by contrast with FBM) velocity
 $v(t)=\dot x(t)$. We consider it in this paper as a basic process, which is ergodic in the absence of an
 external potential. However, when a trapping (e.g., bistable) potential is present, the limit of
 sufficiently large $t'\to \infty$ must be considered. Otherwise, both averages can never coincide even
 for perfectly ergodic processes, requiring some relaxation time to reach a stationary limit within a
 potential trapping domain. This is also the reason why a mean-ergodic, or even correlation-ergodic
 process can be profoundly non-ergodic, e.g., from the viewpoint of statistics of transitions between two
 trapping subdomains \cite{PCCP17,GoychukPRE19,GoychukPRL19Reply}. 
 different from one expected from the relaxation dynamics in such trapping domains.  In such a case,
 swift transitions between two trapping domains can occur in the background of a slow relaxation dynamics
 within the \textit{whole} domain of bistability \cite{GoychukPRE09,GoychukACP12,GoychukPRE19}. Here, the
 main preconditions of the rate theory \cite{HanggiRevModPhys} are violated, and, nevertheless, it can
 remain useful and predictive, to a certain extent \cite{GoychukPRE09,GoychukACP12,GoychukPRE15}. In
 periodic potentials, such an inertial generalization of FBM is also asymptotically ergodic
 \cite{GoychukPRE09}. However, the pace of establishing this ergodic asymptotics heavily depends on the
 potential amplitude in units of the thermal energy $k_BT$, and this asymptotics is not necessarily
 achievable in practice \cite{GoychukPRE09}. These remarks already explain why viscoelastic subdiffusion
 in complex environments is generally not ergodic, and a scatter of single-trajectory averages does not
 contradict the viscoelastic nature of subdiffusion in complex inhomogenous fluids, including cytosol of
 biological cells.

On the practical level, we neither ever have infinitely long trajectories nor infinitely large ensembles.
Hence, both averages never coincide empirically exactly even for ergodic process. However, for a
genuinely ergodic process, they do agree well for sufficiently large $\Delta {\cal T}_w \gg t$ and $M\gg
1$, which is the case of FBM and potential-free GLE subdiffusion \cite{Deng09,
GoychukPRE09,GoychukACP12,MetzlerPCCP}. For this, however, $\Delta {\cal T}_w$ must exceed $t$ by a
factor of at least 100. Notice that some authors consider even ${\cal T}_w \sim t$ or $\Delta {\cal T}_w
\sim 0$ in discussion of ergodic properties \cite{MetzlerPCCP}. It makes, strictly speaking, a little
sense in this context. This point is crucial because most of the existing thus far experimental
confirmations of aging in anomalous diffusion in living cells have ${\cal T}_w$, which at best is only
one hundred times larger than the corresponding $t$. Such proofs lack a profound statistical
significance, especially considering that also $M$ is, at best, several hundred in most experiments, as
we discuss below in the context of our model.

\section{Results and Discussion}

\subsection{Ensemble averages}

\begin{figure}[h]
\vspace{0.8cm}
	\centering
	\includegraphics[height=5.4cm]{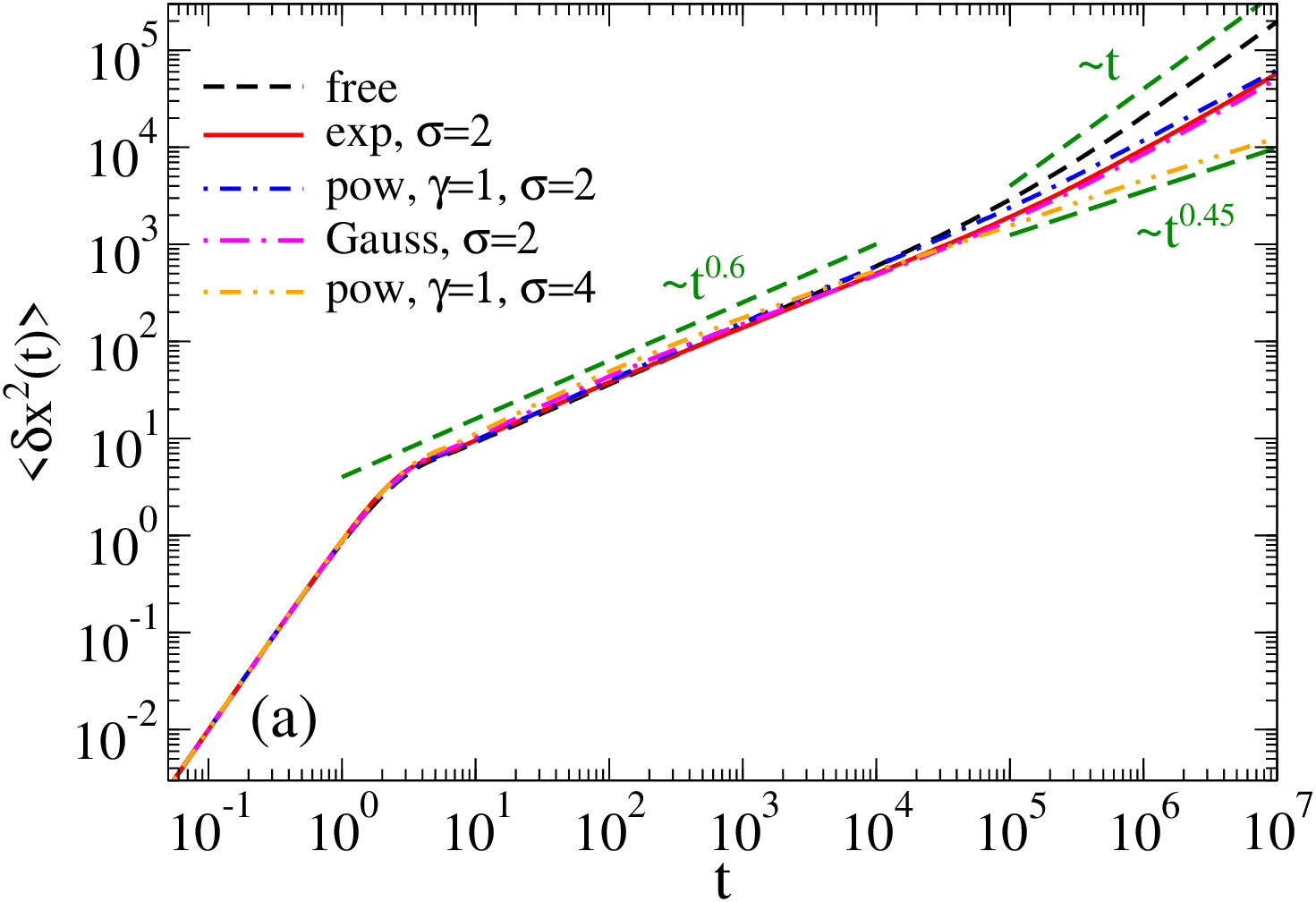} \hfill
	\includegraphics[height=5.4cm]{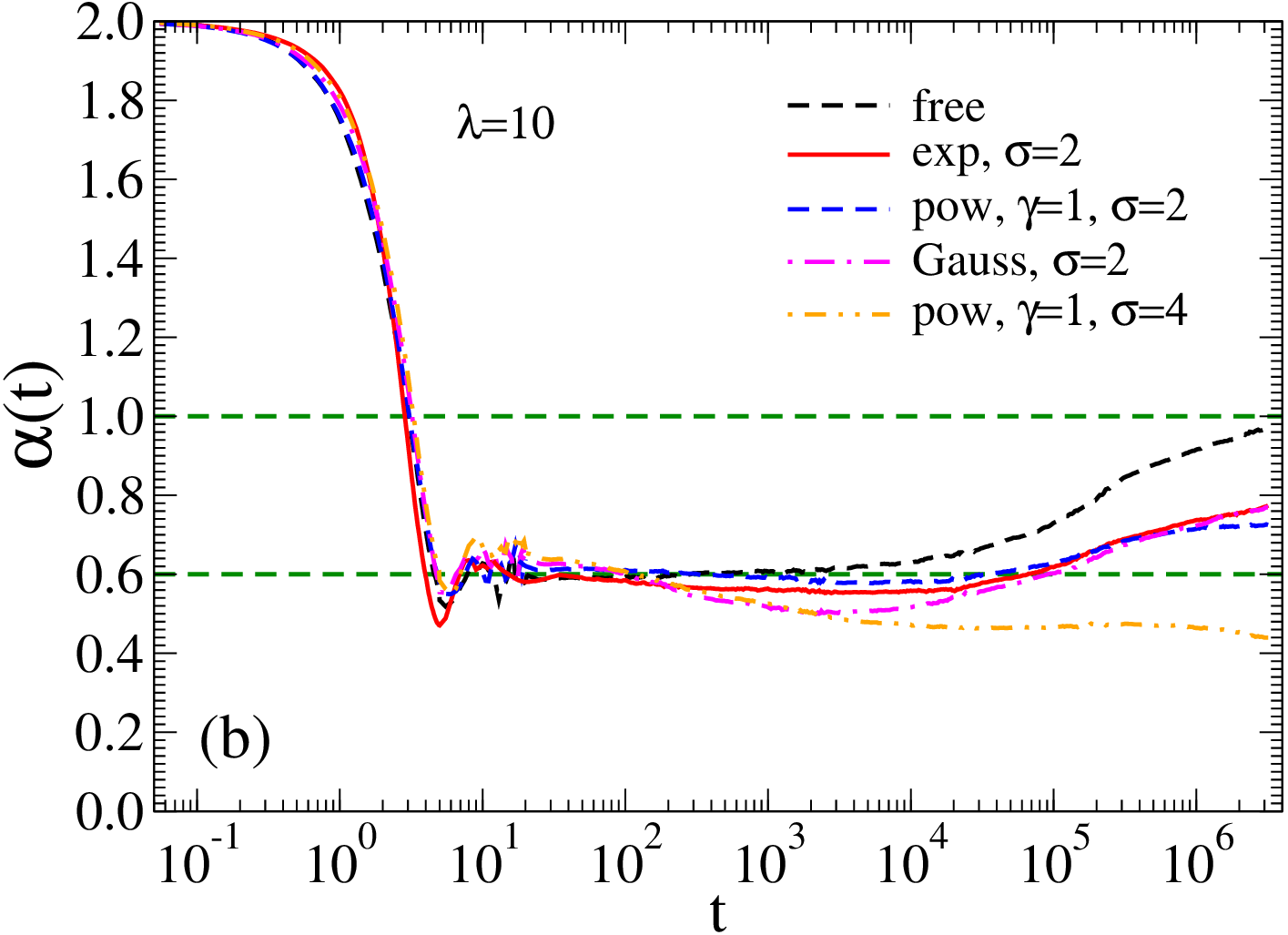}
	\caption{(a) Ensemble-averaged mean-square displacement \textit{vs.} time for several models of
correlation decay (exponential, power-law with $\gamma=1$, and Gaussian) at $\lambda=10$ and several
values of disorder strength $\sigma$ shown in the plot. (b) Dependence of power-law exponent $\alpha(t)$
on time corresponding to (a).}
	\label{Fig2}       
\end{figure}

\begin{figure}[h]
\vspace{0.8cm}
	\centering
	\includegraphics[height=5.4cm]{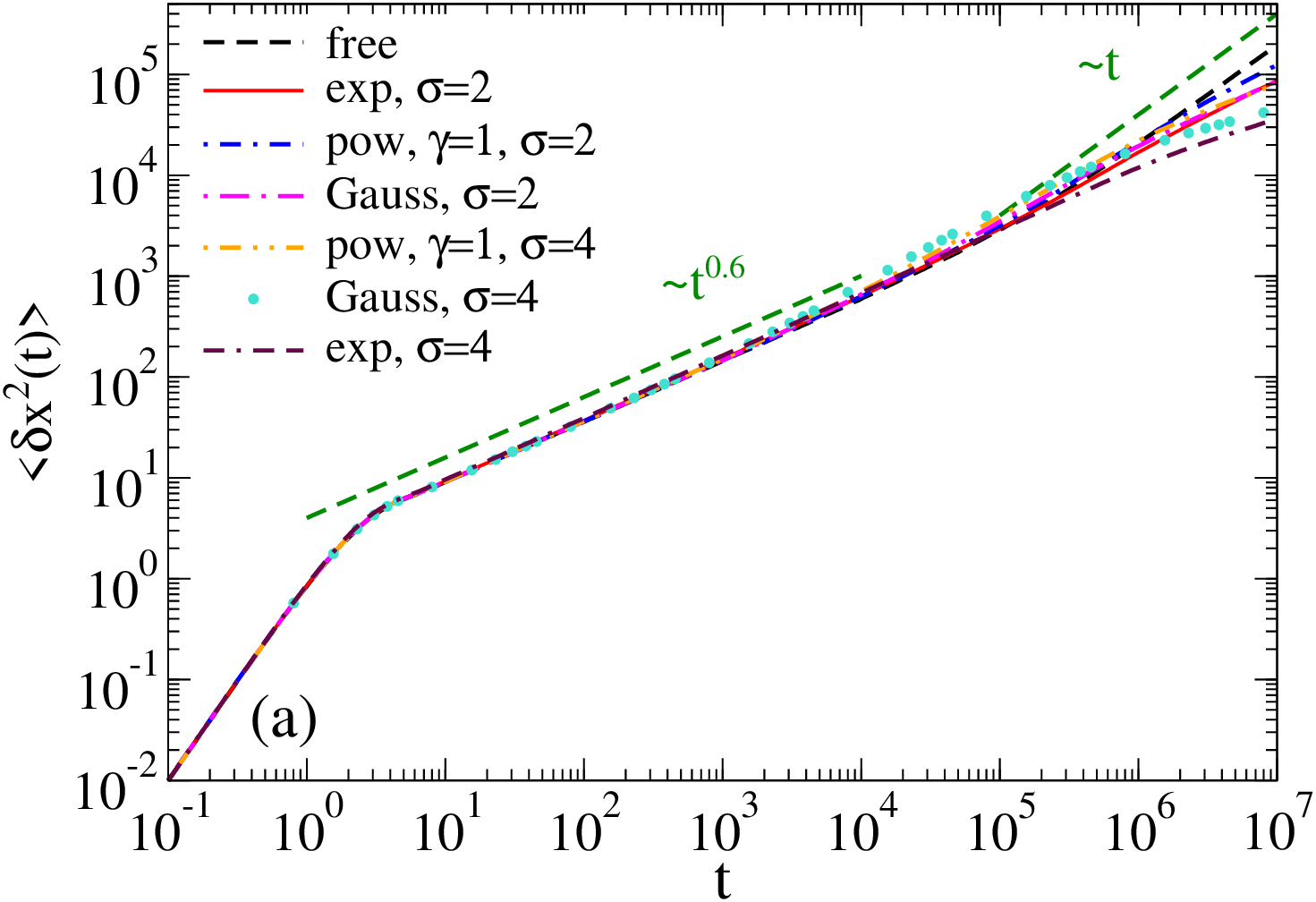} \hfill
	\includegraphics[height=5.4cm]{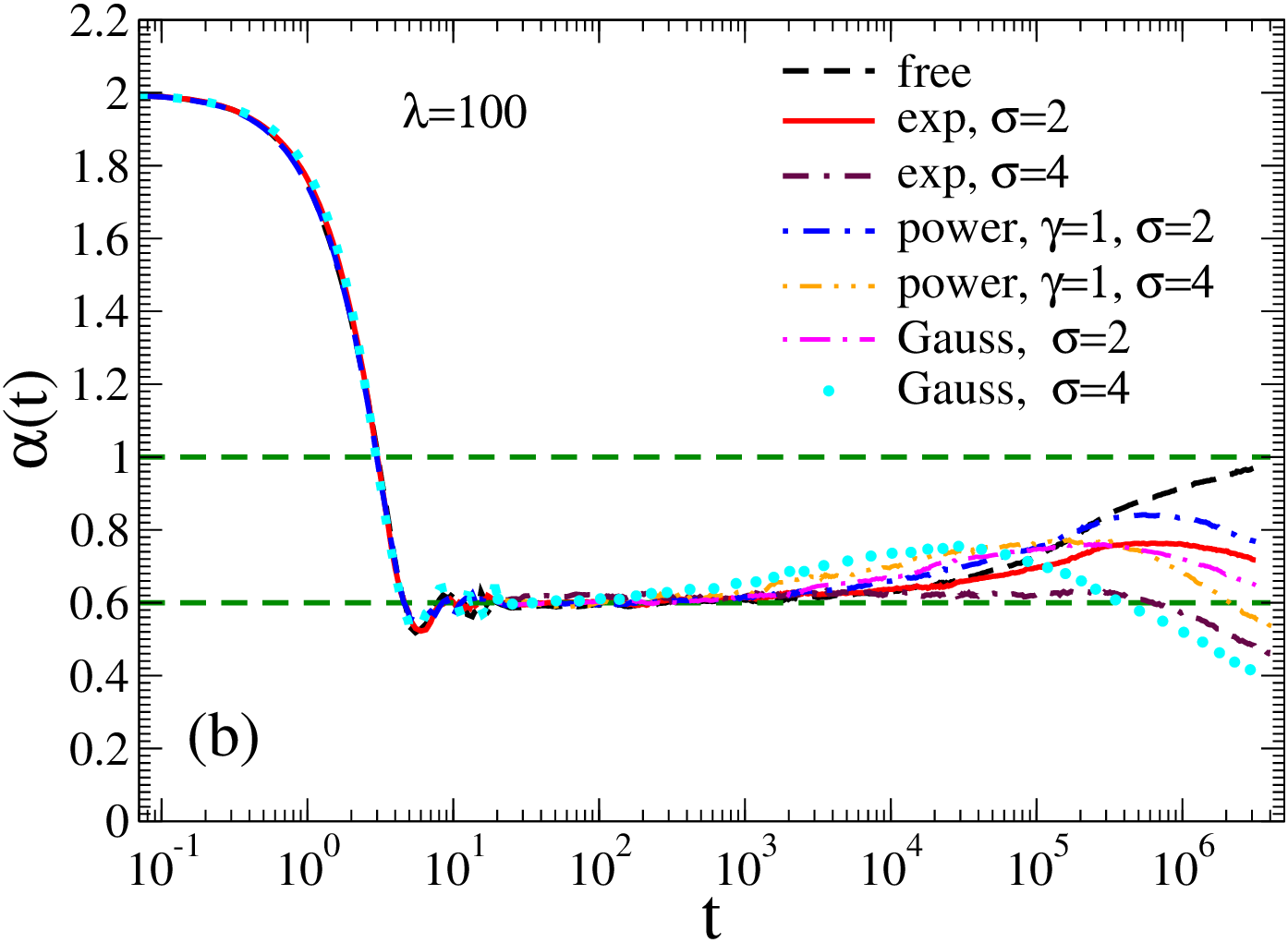}
	\caption{(a) Ensemble-averaged mean-square displacement \textit{vs.} time for several models of
correlation decay (exponential, power-law with $\gamma=1$, and Gaussian) at $\lambda=100$ and several
values of disorder strength $\sigma$ shown in the plot. (b) Dependence of power-law exponent $\alpha(t)$
on time corresponding to (a).}
	\label{Fig3}       
\end{figure}

We first performed calculation of the ensemble averages with $M=10^5$ particles in each case (cf.
\ref{appendA} for pertinent details) using three models of correlation decay and two values of
$\lambda=10$ and $\lambda=100$. The memory kernel approximation is chosen  in such a way that  FLE
subdifussion with $\alpha=0.6$ should last in the absence of potential until about $t=10^3$. Then, it
gradually changes into  asymptotically normal diffusion, see the cases of free diffusion in Figs.
\ref{Fig2}, \ref{Fig3} (a). This behavior is convenient to characterize by a time-dependent power law
exponent $\alpha(t)=d\ln \langle \delta x^2(t)\rangle/d\ln t$, see in Figs. \ref{Fig2}, \ref{Fig3} (b). 
Initially  it is always ballistic, $\alpha=2$, $ \langle \delta x^2(t)\rangle\approx (v_Tt)^2$, given
initially thermally distributed velocity variable in this work. Then, after some transient an FBM regime
with $\alpha_{\rm eff }\approx 0.6$ establishes for $t\geq 10$ and lasts until $t\sim 10^3- 10^4$. Then,
significant deviations from $\alpha=0.6$ start. The power exponent grows and the regime of normal
diffusion gradually establishes until the end of simulations at $t_{\rm max}=10^7$. This behavior
strongly reminds one for diffusion of lipids  in 1,2-Distearoyl-sn-glycero-3-phosphocholine (DSPC),
1-stearoyl-2-oleoyl-sn-glycero-3-phosphocholine (SOPC), and 1,2-Dioleoyl-sn-glycero-3-phosphocholine
(DOPC) lipid bilayers  in \cite{JeonPRL12,Metzler16}, see Fig. 8 in \cite{Metzler16}. Therein, FBM and
FLE subdiffusive regime with $\alpha=0.6$ lasts until about 10 ns and then changes gradually into normal
diffusion. With $\tau_v=0.323$ ps (\ref{appendB}), our $t=10^4$ corresponds to 3.23 ns and $t=10^7$ to 3.23
$\mu$s. 

The presence of a random potential radically changes this simple picture. Within the time-range of purely
viscoelastic subdiffusion ($t<10^3$) disorder practically does not influence it on the ensemble level,
see in Figs. \ref{Fig2}, \ref{Fig3}, (a). It is especially surprising for a very rough exponentially
correlated disorder with $\lambda=10$ and $\sigma=2$ in Fig. \ref{Fig2} (a) (see a typical realization of
such a disorder in Fig. \ref{Fig1} (a)), and also for a power-law correlated disorder for $\sigma=4$ and
$\lambda=10$. Naive expectation based on the de Gennes-B\"assler-Zwanzig renormalization is that the
diffusional spread should be suppressed by the factor $\exp[-\sigma^2/(k_BT)^2]$, which is approximately
$0.01832$ for $\sigma=2$ and $1.12535\times 10^{-7}$ for $\sigma=4$. However, this expectation is not
justified; see in Fig. \ref{Fig2} (a). Time-dependent $\alpha(t)$ shows some dynamics in Fig. \ref{Fig2}
(b). However, changes on the level of $\langle \delta x^2(t)\rangle $ are not easy to detect in part (a).
This striking feature was described earlier within a bit different model (no inertial effects, diffusion
is initially normal due to a normal friction component in GLE) in \cite{PCCP18}. Also, in periodic
potentials, GLE subdiffusion is asymptotically insensitive to the presence of potential
\cite{GoychukPRE09, GoychukACP12}. For a sinusoidal potential, it can be even proven rigorously within a
quantum-mechanical setting while considering the sub-Ohmic model of generalized Brownian motion
\cite{WeissBook}, which corresponds to the considered viscoelastic GLE subdiffusion \cite{GoychukACP12}. In periodic
potentials, however, transient effects can last very long, depending on the potential amplitude
\cite{GoychukPRE09, GoychukACP12}. In the present case, no related transients can be seen on the ensemble
level, even for $\sigma=4$ in Fig. \ref{Fig2} (a), which is quite surprising. The influence on the level
of single-trajectory averages is, however, in the case $\lambda=10$, profound, see below. Next,
transition to normal diffusion regime does not occur until $t_{\rm max}=10^7$.  A new anomalous diffusion
regime caused by disorder is gradually established for $t>10^5$, with vibrant transient features, which
depend on the type of correlation decay, $\sigma$, and $\lambda$.

Indeed, for exponential correlations and $\sigma=2$,  $\alpha(t)\approx 0.55-0.57<0.6$ for $t=10^2-10^4$
in Fig. \ref{Fig2} (b) (full red line therein). For $t>10^4$ it starts to grow slowly and reaches
$\alpha\approx 0.774$ at $t=3.2\times 10^6$. In this respect, it is worth noting that in the case of
overdamped normal diffusion in such a potential, $\alpha(t)$ reaches a minimal value $\alpha\approx 0.70$
\cite{GoychukPRL14,Goychuk2017} before it starts to grow logarithmically  slow in time until unity. We
expect such a slow growth also in the case under study. However, the asymptotic regime of normal
diffusion is numerically out of reach.  Also in the case of Gaussian correlation decay, $\alpha(t)$ first
drops slightly below $0.6$, reaching its minimum $\alpha_{\rm min}\approx 0.50$ at $t\approx 2\times
10^3$ and then slowly grows until $\alpha\approx 0.77$ at $t=3.2\times 10^6$. Especially interesting is
the case of power-law correlations with divergent correlation length because in this case subdiffusion
caused by the correlated Gaussian disorder alone (without viscoelastic memory effects) is the longest
\cite{Goychuk2017}. In this case, $\alpha(t)$ stays close to $0.6$ for $t$ between $10$ and $10^5$. Then,
it grows and stays nearly constant $\alpha(t)\approx 0.72$ from $t=10^6$ till the end of the simulations.
In this respect, $\alpha\approx 0.70$ is about the minimal value of transient subdiffusion caused also by
Gaussian power-law correlated disorder \cite{Goychuk2017}.  In the case of normal overdamped diffusion in
considered random potentials, the spatial range of nonergodic transient subdiffusion is determined by a
nonergodicity length $L_{\rm erg}$, which is a typical length scale, where the statistical weight
function $w(x)=\exp[-U(x)/(k_BT)]$ lacks self-averaging \cite{GoychukPRL14,Goychuk2017}. For $\sigma=2$,
$L_{\rm erg}\approx 35 \lambda$ in the case of exponential correlations, $L_{\rm erg}\approx 52 \lambda$
in the case of Gauss-decaying correlations, and $L_{\rm erg}\approx 167 \lambda$ in the case of power-law
decaying correlations with $\gamma=0.8$, see Table I in \cite{Goychuk2017}. Hence, in the case of 
power-law correlations (here $\gamma=1$ \textit{vs.} 0.8 in \cite{Goychuk2017}) one can expect that
subdiffusion  will last at least until about $\delta x\sim 1.67\times 10^3 x_0$ for $\lambda=10$ and
$\delta x\sim 1.67\times 10^4 x_0$ for $\lambda=100$, which for $x_0=0.026$ nm (\ref{appendB}) would
correspond to 43.42 nm and 434.2 nm, correspondingly. With increasing $\sigma$, $L_{\rm erg}$ grows
super-exponentially fast. Already for $\sigma=4$, $L_{\rm erg}\sim 10^6\lambda$ for a short-range
correlated disorder \cite{Goychuk2017}. Then, the corresponding nonergodicity length reaches practically
a macroscale, $L_{\rm erg}\approx 0.26$ mm  and $2.6$ mm at $\lambda=10$ and $\lambda=100$,
correspondingly. In this respect, the case of power law correlation in Fig. \ref{Fig2} with $\sigma=4$ is
interesting. From $t=10^4$ on in part (b), $\alpha(t)\sim 0.44-0.47 $ with $\alpha_{\rm min}\approx 0.44$
at the end of simulations, where it seems to slightly decline further. It agrees well with $\alpha_{\rm
min}\approx 0.43$ for power-law correlations in Fig. 3 (b) of Ref. \cite{Goychuk2017} obtained for
memoryless diffusion.

The case of $\lambda=100$ in Fig. \ref{Fig3} is interesting for some lipid-proteins systems. For such a
large $\lambda$ and a smooth disorder, the time range of viscoelastic diffusion is restricted virtually
by motion within one local minimum of the random potential, cf. Fig. \ref{Fig1} (b), of a typical spatial
size $\lambda$. In this case, $\alpha(t)$ dynamics in Fig. \ref{Fig3} (b) shows an interesting novel
non-monotonic feature in all studied cases. Namely, it first starts to grow, like in the potential-free
case. However, after a maximum is reached, $\alpha(t)$ drops. For example, for exponential correlations
(in this singular case, a very rough multiple-extrema structure of potential holds also on a spatial
scale much smaller than $\lambda$) and $\sigma=2$, $\alpha_{\rm max}(t)\approx 0.763$ is reached at
$t=5.4\times 10^5$. After this, it drops to $\alpha(t)\approx 0.717$ at $t=3.17\times 10^6$. Given the
results in \cite{Goychuk2017}, one can predict that it will further drop to $\alpha\approx 0.70$ and then
start to grow logarithmically slow to the asymptotically normal value. Furthermore, for Gauss-decaying
correlations and $\sigma=2$, $\alpha_{\rm max}(t)\approx 0.760$ is reached at $t=2.84\times 10^5$, and it
drops to $\alpha(t)\approx 0.648$ at $t=3.17\times 10^6$. For this model of the disorder, $\alpha_{\rm
min}\approx 0.6$, see in Fig. 3 (d) in \cite{Goychuk2017}. Hence, it is expected that with increasing
time, it will drop further to this minimum value and then gradually increase to the asymptotically normal
value. In case of power-law correlations and $\sigma=2$, $\alpha_{\rm max}(t)\approx 0.842$ is reached at
$t=5.57\times 10^5$, and it drops to $\alpha(t)\approx 0.768$ at $t=3.17\times 10^6$. One expects that it
will further drop to $\alpha_{\rm min}=0.70$ and then gradually grow. Interestingly, a reminiscent
nonmonotonous behavior of $\alpha(t)$ was found both for lipids and proteins in MD simulations of
disordered lipid-protein systems, see Figs. 15, 16 in \cite{Metzler16}, and references therein. 

Next, for $\sigma=4$ and exponential correlations, $\alpha_{\rm min}\approx 0.375$ in the case of normal
diffusion \cite{GoychukPRL14} (see inset in Fig. 1 therein) and it can stay for a very long time nearly
constant. In Fig. \ref{Fig3} (b), the minimal value reached at $t=3.17\times 10^6$ for exponential
correlations is $\alpha\approx 0.456$. The true minimum is hence still not reached. For power-law
correlations in this figure, the minimum at the end of simulations is $\alpha\approx 0.529$. It is well
above the expected minimal $\alpha_{\rm min}\approx 0.43$, which is still not reached. Also, for Gaussian
correlations,  $\alpha\approx 0.406$ at the end is still above the expected minimum. After it will be
reached, $\alpha(t)$ is expected to stay nearly constant for a long time and then logarithmically slow
grow until unity. These predicted features are, however, beyond any numerical study currently possible. 

It must be emphasized that in our estimations we tacitly assume that on the time scale exceeding the
long-time cutoff of the memory kernel $\tau_h$ we can think in terms of normal diffusion with an
effective friction coefficient $\eta_{\rm eff}=\int_0^{\infty}\eta(t)dt$, which takes place in a random
potential. The validity of this assumption should be checked in the subsequent research. One cannot
exclude that a synergy of viscoelastic memory effects and nonergodic effects due to the randomness of
potential will tremendously increase nonergodicity length.

\subsection{Distribution of particle positions}

A very intriguing question emerges on the origin of the exponential power distribution
\begin{eqnarray}
P(x,t)\propto \exp\left (-|x/x_{\chi}(t)|^{\chi(t)}\right )\;.
\label{exp-pow}
\end{eqnarray}
 Such a distribution is commonly found in viscoelastic biological subdiffusion, e.g., for RNA-protein
complexes in various cells with a fixed power exponent $\chi=1$ in Ref. \cite{Lampo17}. Furthermore,
crowded protein-lipid systems also display this distribution with $\chi=1.4-1.6$
\cite{JeonPRX,Metzler16}. Very important in this respect is that normal diffusion in disordered Gaussian
potentials typically displays this distribution with a time-dependent $\chi(t)$, see Fig. 2 of Supplemental
Material in Ref. \cite{GoychukPRL14}, where $\chi\sim 1.4-1.6$ for $\sigma=2-4$. Hence, it is expected
that the viscoelastic subdiffusion of a restricted temporal range in Gaussian potentials will display
this striking feature.

\begin{figure}[h]
\vspace{0.8cm}
	\centering
	\includegraphics[height=5.4cm]{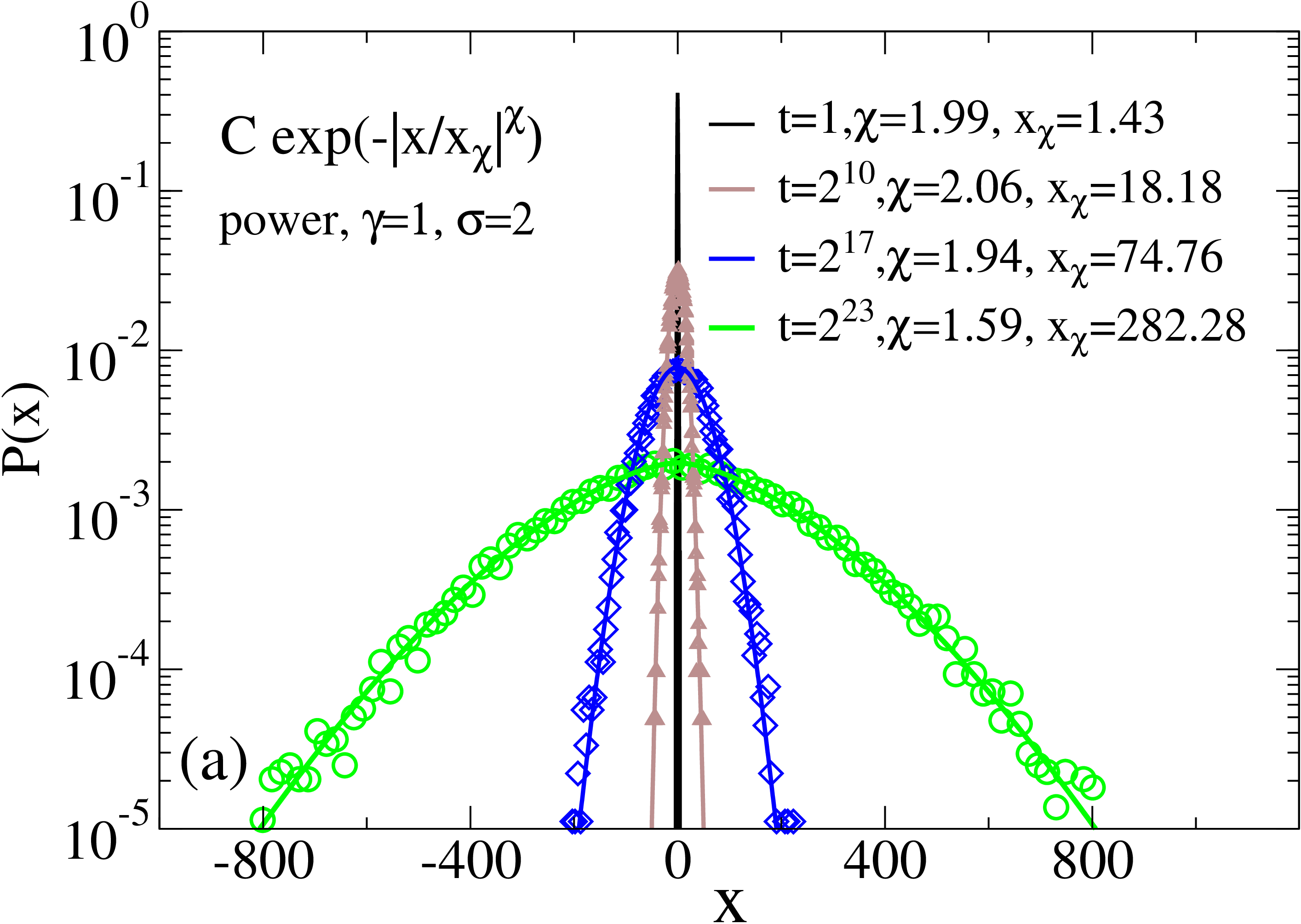} \hfill
	\includegraphics[height=5.4cm]{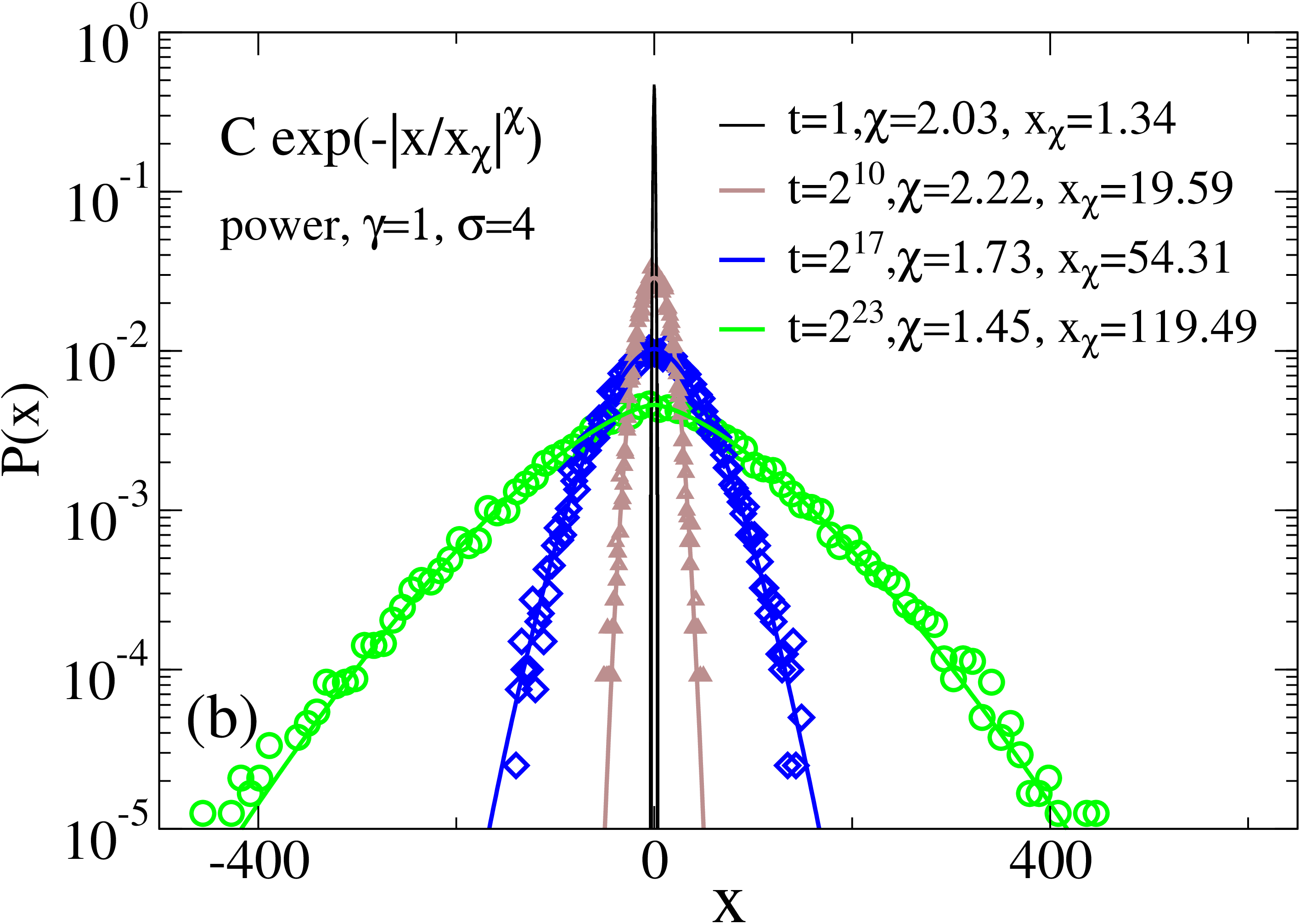}
	\vfill \vspace{1cm}
		\includegraphics[height=5.4cm]{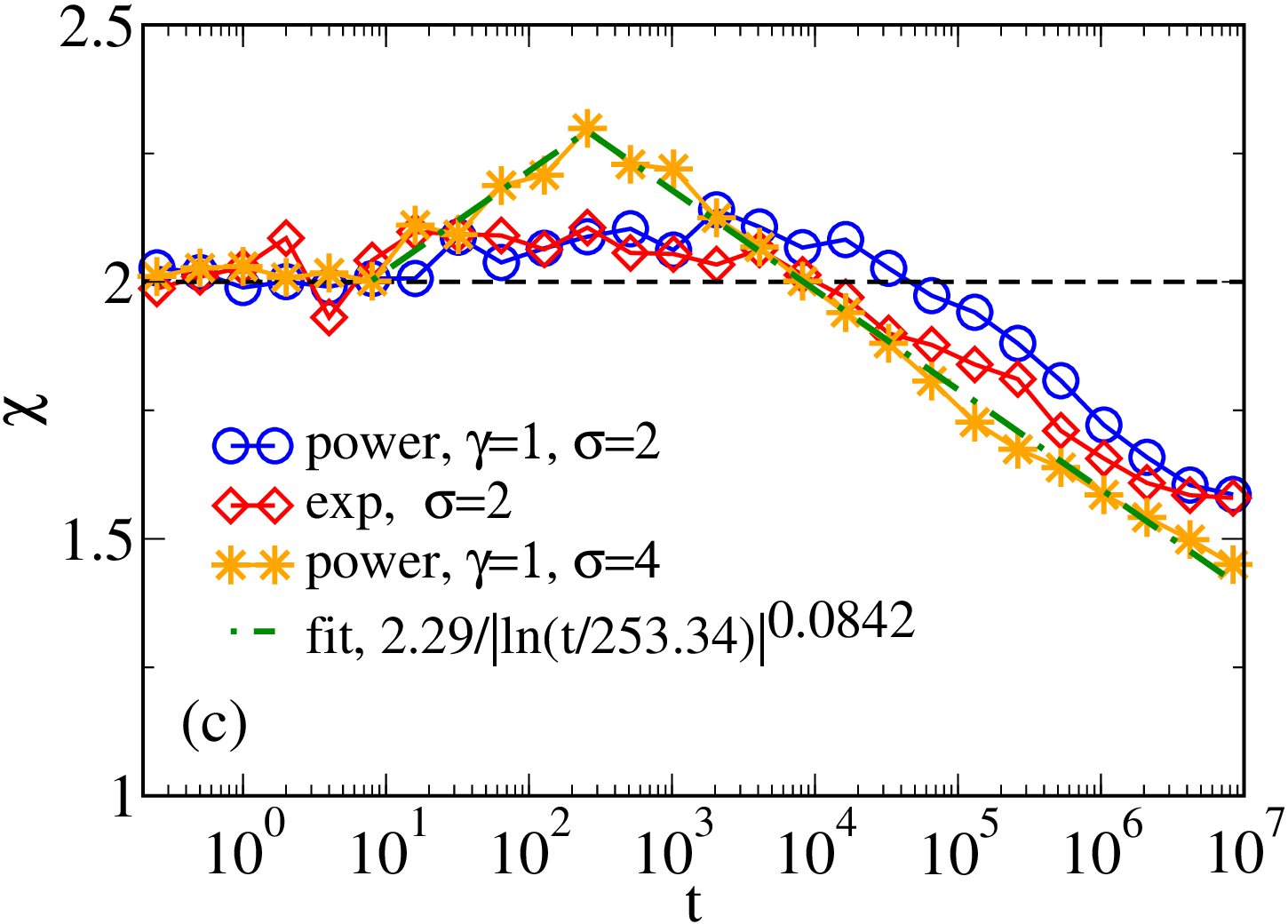} 
	
	\caption{Distribution of particle positions for several values of time in the case of power-law correlations with $\gamma=1$, $\lambda=10$ and (a) $\sigma=2$, (b) $\sigma=4$. Symbols display the numerical results and lines the corresponding fits with Eq. (\ref{exp-pow}) and parameters shown in plots. Panel (c) displays time-dependence $\chi(t)$ for several models of correlations with $\lambda=10$, and a fit with (\ref{chi_relax}) in one case.}
	\label{Fig4}       
\end{figure}

Indeed, our simulations in Fig. \ref{Fig4} confirm this expectation. They reveal the following common
features. Initial delta-distribution spreads first approximately as a Gaussian, $\chi=2$, see in panel (c). Then,
during the viscoelastic stage, $\chi$ fluctuates around $\chi=2$, and can even reach $\chi\approx 2.3$,
see the case of power-law correlations in panel (c). It drops below $\chi=2$, beyond the temporal range
of viscoelastic correlations, where the very occurrence of subdiffusion is on the count of the randomness
of potential. For $\sigma=2$, $\chi(t)$ relaxes down to $\chi_{\rm min}\approx 1.59$ for several models
of correlations under study, see in panel (a) for power-law correlations. For $\sigma=4$, the minimal
value of $\chi$ in parts (b), (c) is $\chi=1.45$. However, the real minimum, in this case, is still not
achieved until the end of the simulations. The relaxation law is inverse logarithmic
\begin{eqnarray}\label{chi_relax}
\chi(t)\propto 1/[\ln(t/t_c)^r]
\end{eqnarray}
with $t_c\approx 253.34$ and $r\approx 0.0842$ for $\sigma=4$ in the case of power-law correlations, see
in Fig. \ref{Fig4} (c). After reaching the minimum, it is expected that $\chi$ will stay nearly constant
for a while and then it will gradually increase to $\chi=2$ asymptotically, while the normal diffusion
limit will gradually be achieved. To realize this feature, see the case $\sigma=1$ in the inset of Fig. 2
of Supplemental Material in Ref. \cite{GoychukPRL14}. In the case $\lambda=100$, the relaxation behavior
in Fig. \ref{Fig4} (c) is shifted to the right (not shown). For example, for $\sigma=2$ and power-law
correlations, $\chi\approx 1.736$ at the end of the simulations, and for $\sigma=4$ it arrives at
$\chi\approx 1.704$, i.e., indeed $\chi_{\rm min}$ is still far from being reached in both cases.

\subsection{Trajectory averages}

\begin{figure}[h]
\vspace{0.4cm}
	\centering
	\includegraphics[height=4.5cm]{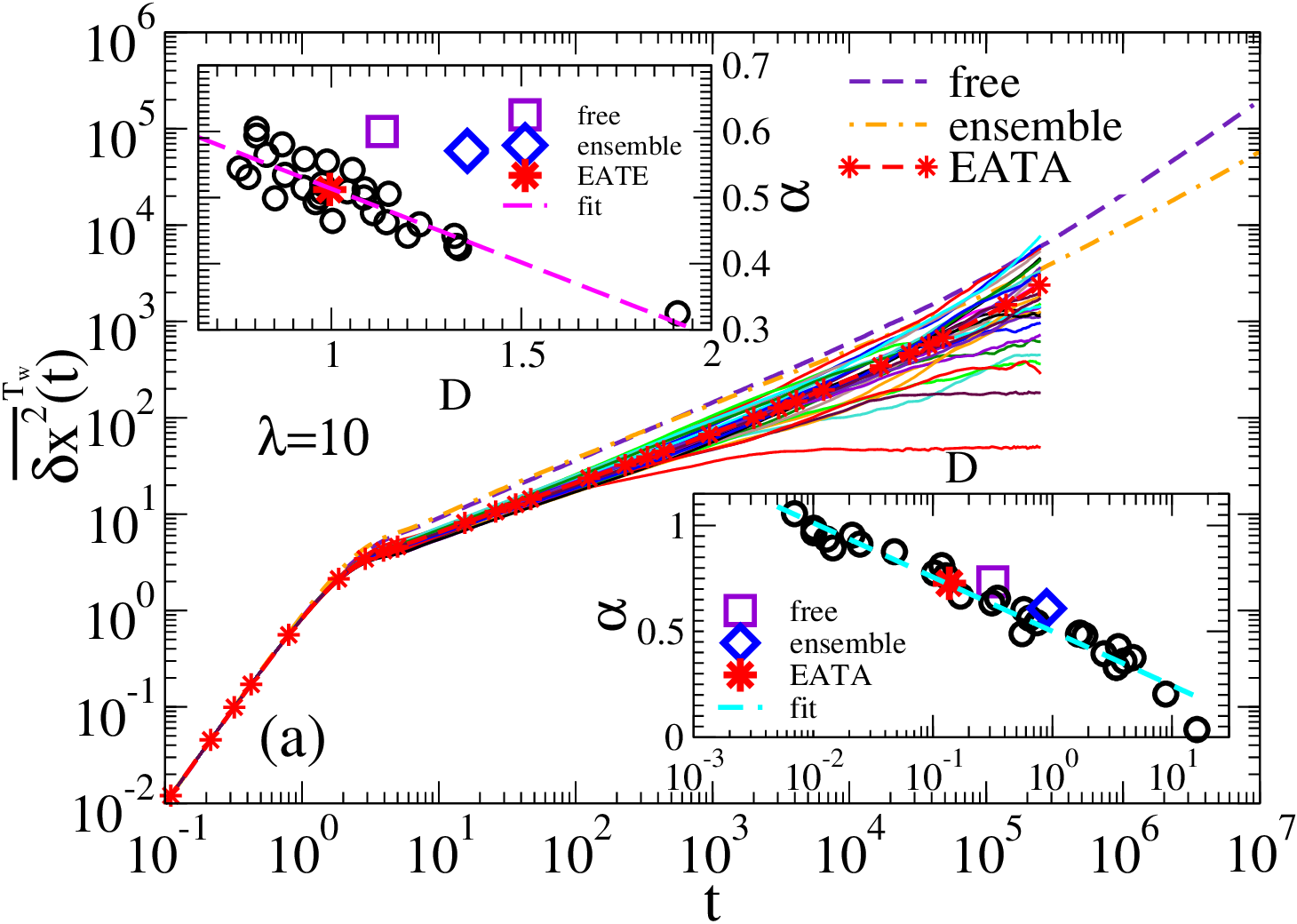} \hfill
	\includegraphics[height=4.5cm]{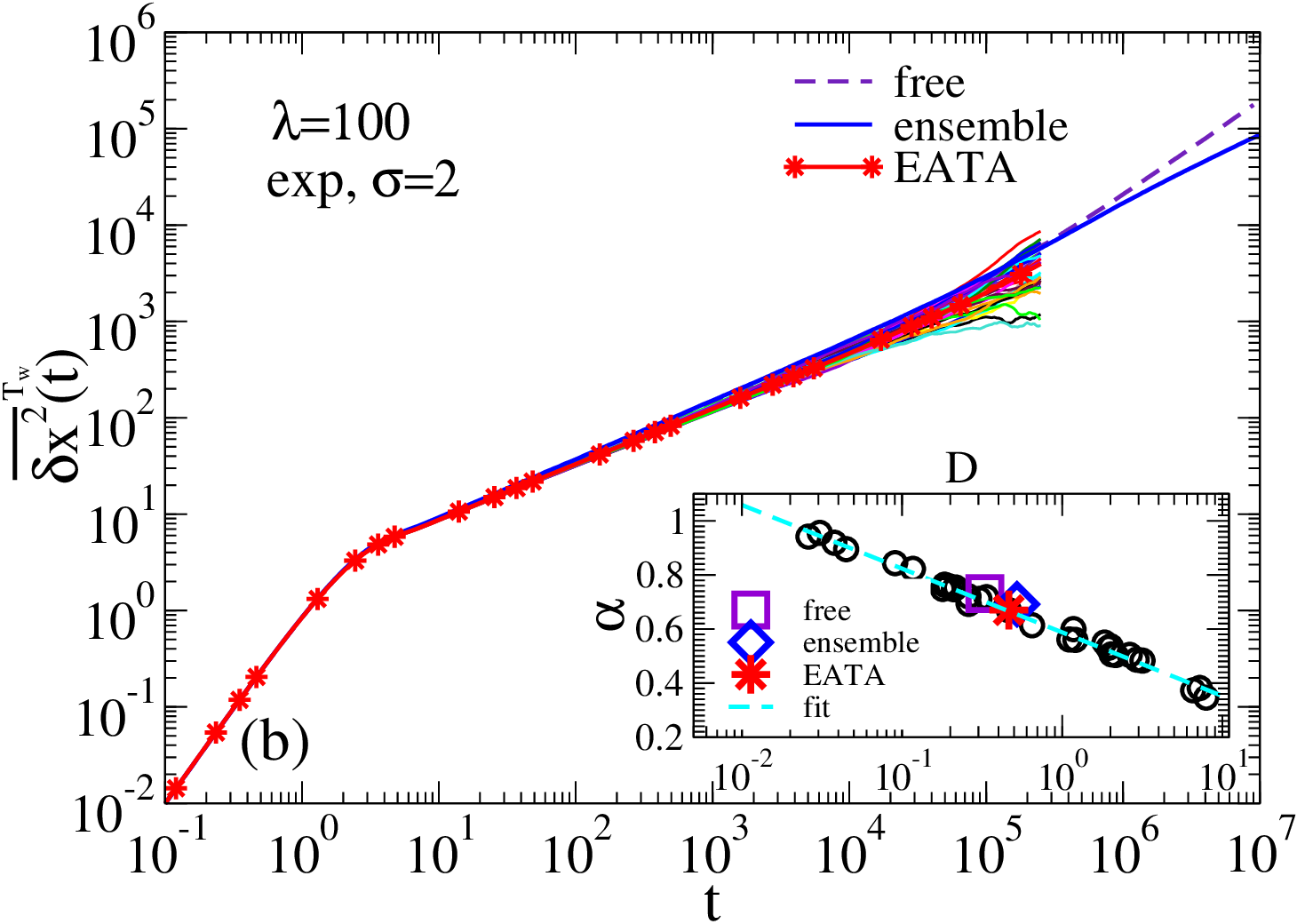} \vfill  \vspace{1cm}
	\includegraphics[height=4.5cm]{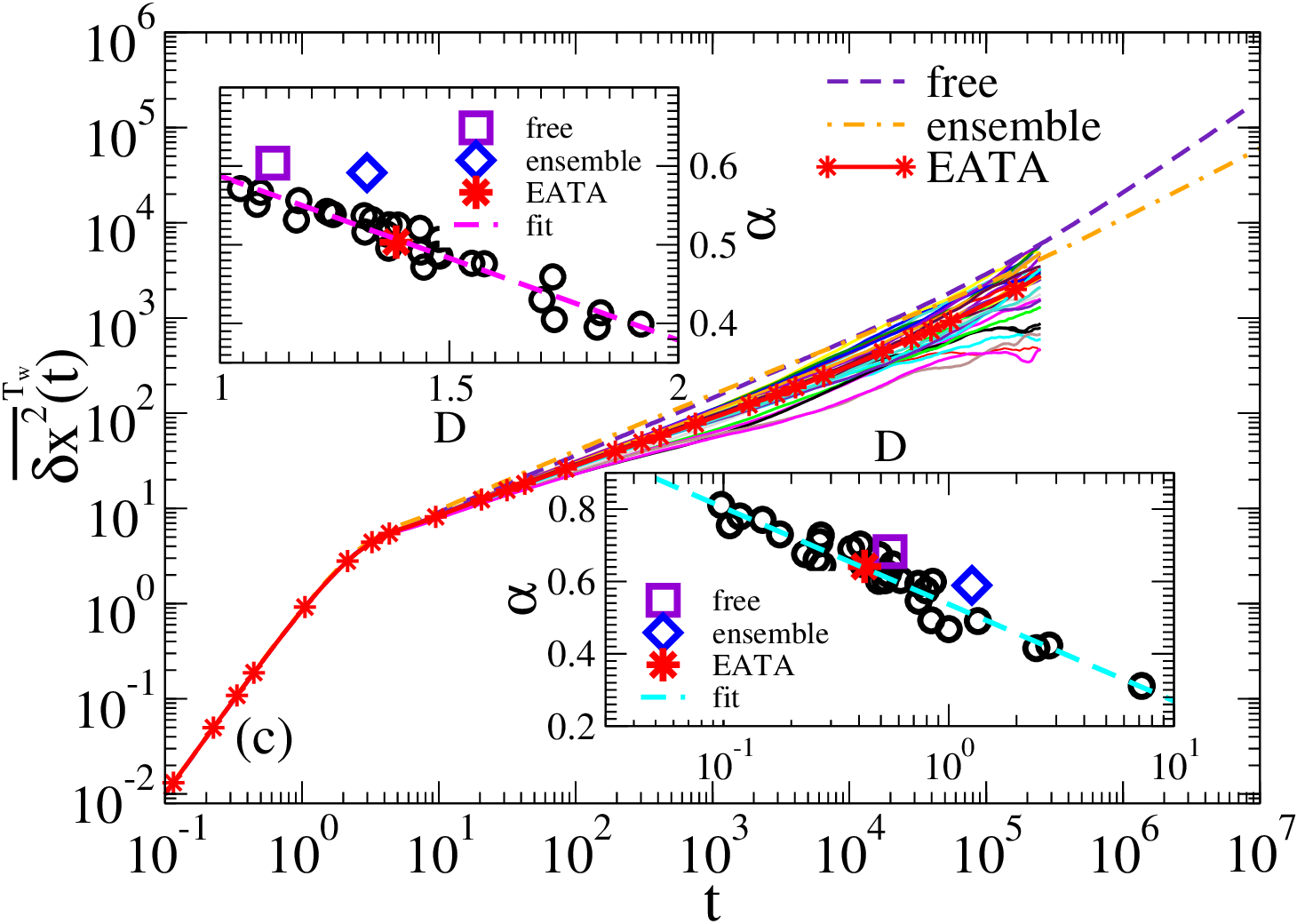} \hfill
	\includegraphics[height=4.5cm]{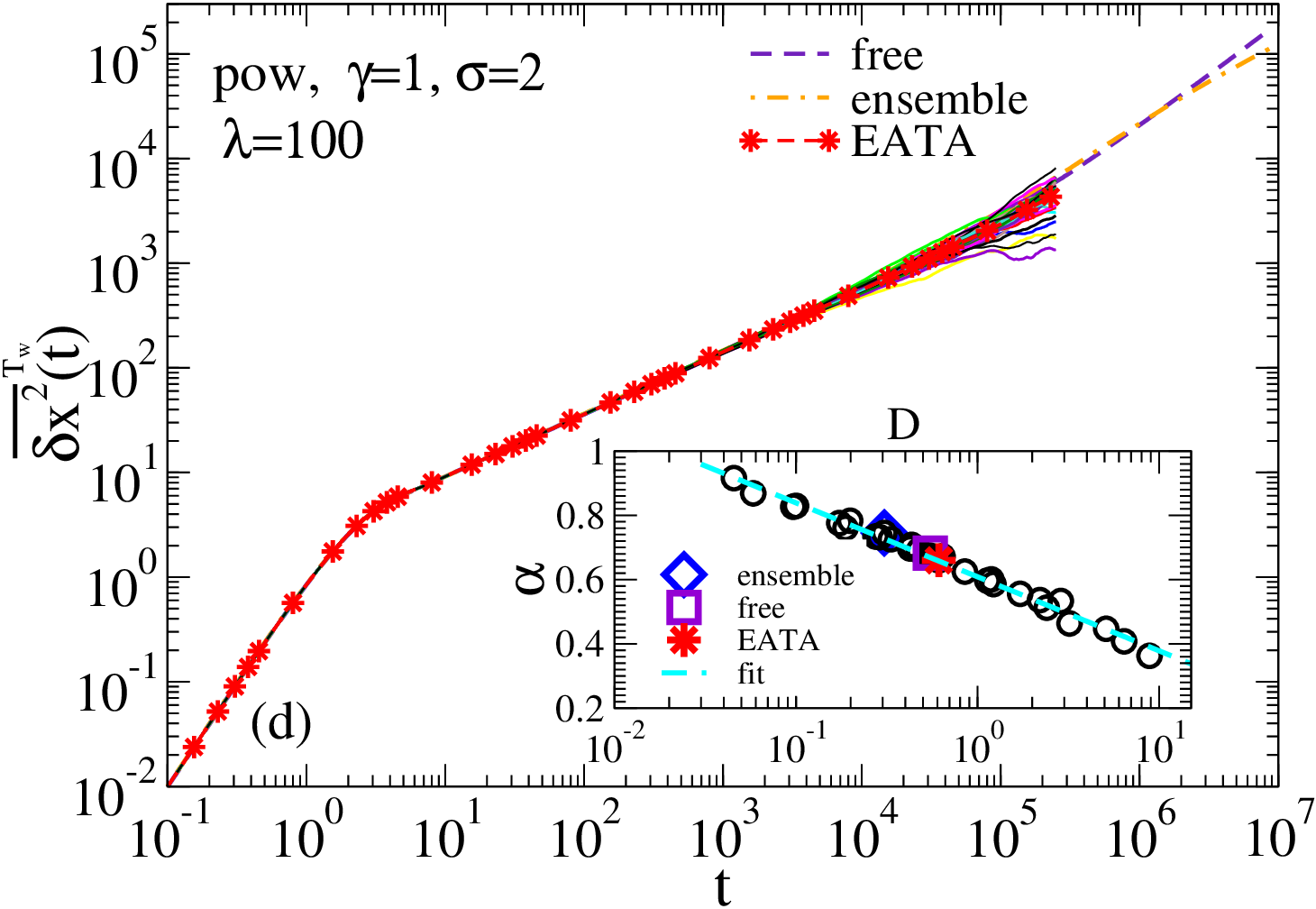} \vfill  \vspace{1cm}
	\includegraphics[height=4.5cm]{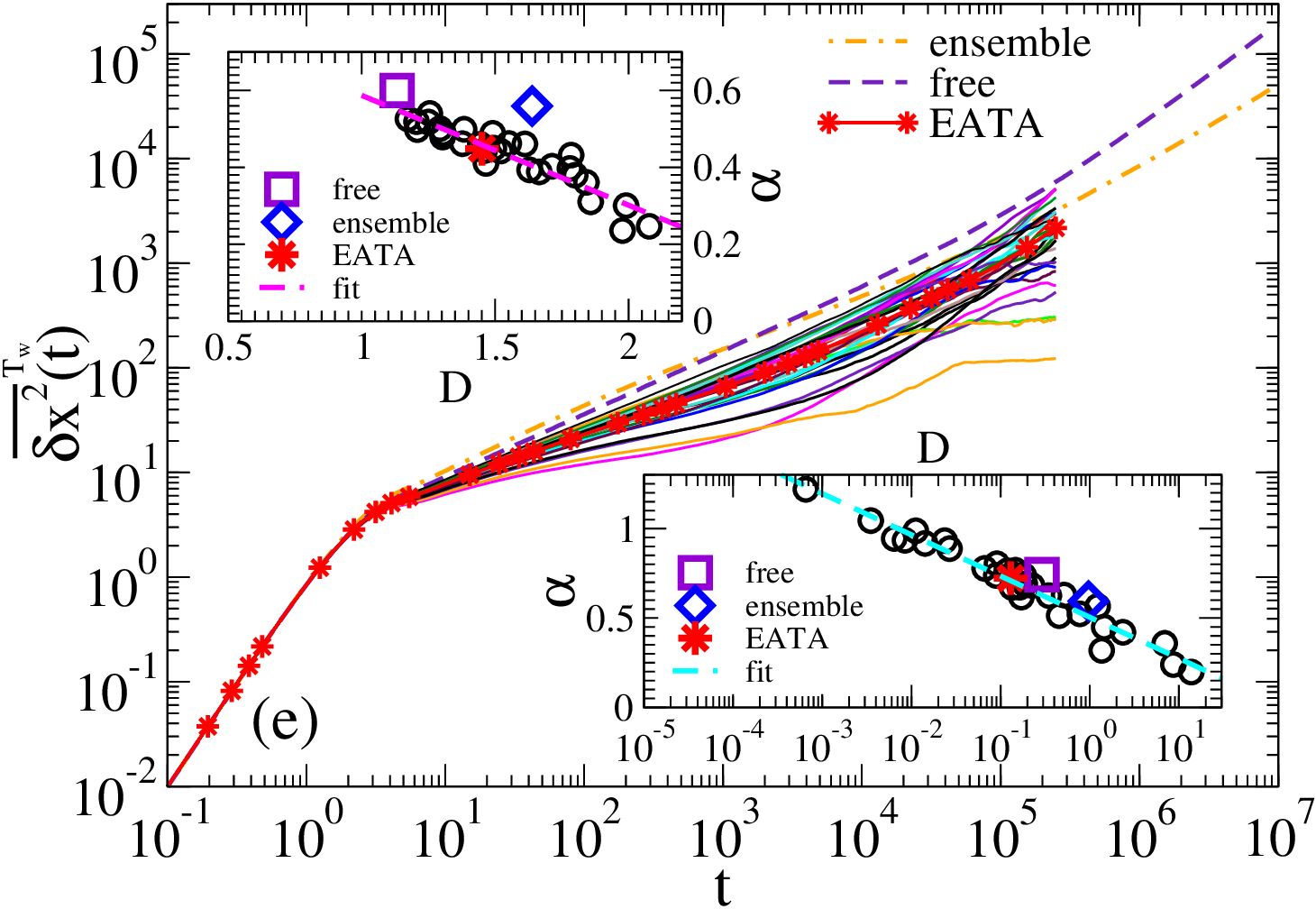} \hfill 
	\includegraphics[height=4.5cm]{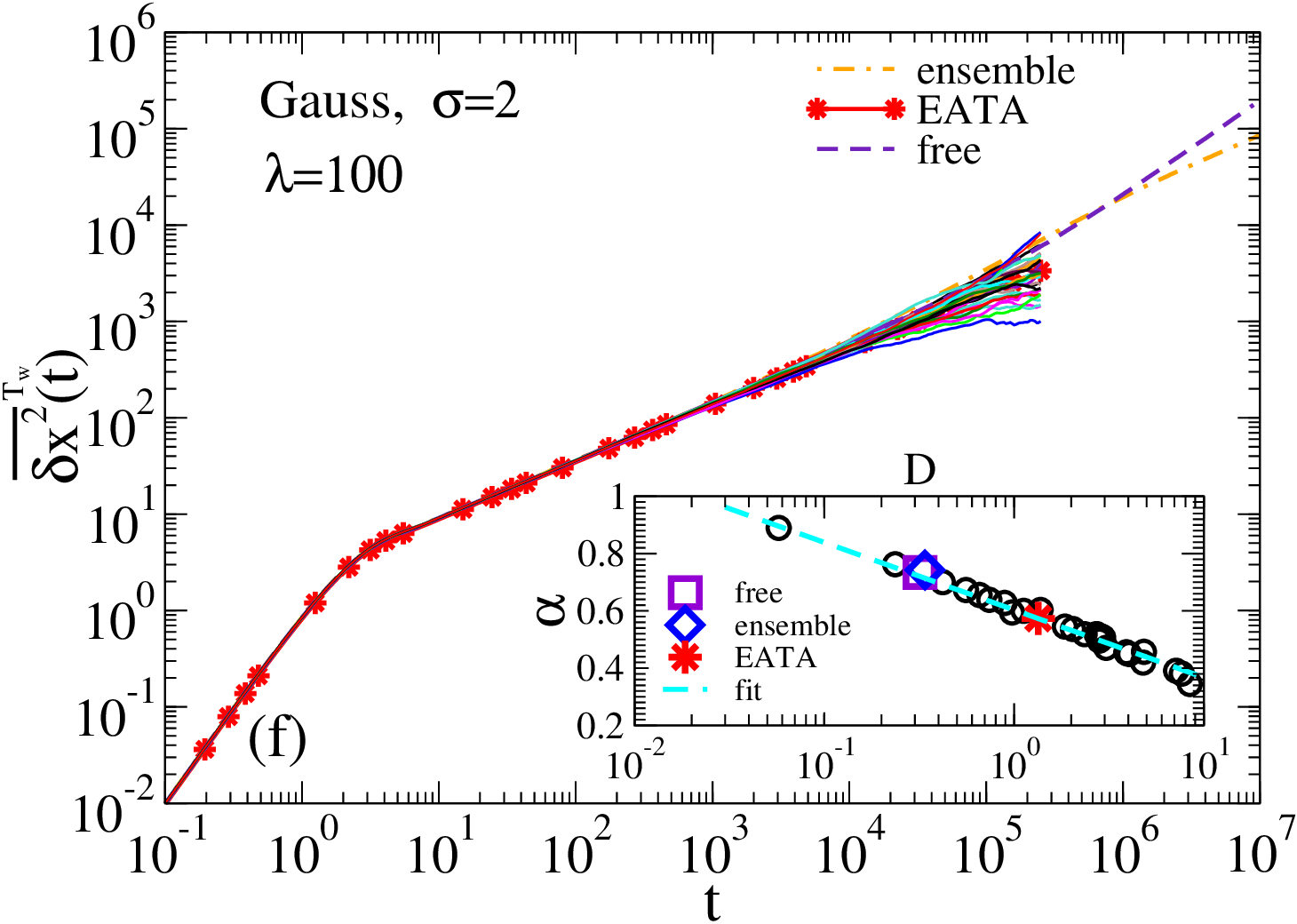}
	\caption{Single-trajectory averages for (a,b) exponential correlations, (c,d) power-law
correlations with $\gamma=1$, (e,f) Gauss-decaying correlations for $\sigma=2$ and two values
$\lambda=10$ (a,c,e) and $\lambda=100$ (b,d,f). In each case, 30 trajectory averages are displayed. They
are to compare with the ensemble averages in the cases of free diffusion and for the specific potential
disorder also shown. Besides, ensemble-averaged time-averages (EATAs), $\left \langle \overline{\delta
x_i^2(t)}^{{\cal T}_w} \right\rangle$ are shown in each case. Upper insets in (a,c,d) show pairs
$(D,\alpha)$ derived from the fits of single-trajectory averages by $2Dt^{\alpha}$ within the
time-interval $[10,10^3]$. Lower insets in (a,c,d) show the corresponding pairs derived from the fits
within the time-interval $[10^3,2\times 10^5]$. In (b,d,f) the corresponsing fits are done within the
time-interval $[10,2\times 10^5]$. Straight lines in upper insets display the best fits of numerical data
with  Eq. (\ref{linear}). Straight lines in lower insets display best fits with  Eq. (\ref{correl}). In
both cases, fitting parameters are given in Table \ref{TableI}. }
	\label{Fig5}       
\end{figure}

Next, we proceed with studying single-trajectory averages, which are of special interest because disorder generally implies a broken ergodicity. For this, we use ${\cal T}_w=2\times 10^6$ and evaluate single-averages until $t_{\rm max}=0.1{\cal T}_w$. The results are shown in Fig. \ref{Fig5} for $\sigma=2$, three models of correlation decay and two values of $\lambda$. Consider first the case $\lambda=10$ in parts (a,c,e). First of all, in the initial regime, $t<2$, (including thermal ballistic diffusion), self-averaging occurs for every trajectory. No scatter is present. In this regime, ensemble and trajectory averages coincide. Thus, diffusion is ergodic. For $t>2$ in the cases of exponential and Gaussian correlations and $t>10$ for power-law correlations, the scatter becomes visible. It is noticeable that it becomes enhanced for $t>10^3$, beyond the range of viscoelastic memory effects. Hence, we consider two different time intervals (i) $[10,10^3]$ and (ii) $t>10^3$ to characterize diffusional properties with a pair of random variables $(D,\alpha)$  for each separate trajectory defined on the corresponding time intervals. The results are depicted in upper and lower insets in parts (a,c,e), correspondingly.  They display a striking feature. Namely, while one of the variables can be considered completely random, another is tightly related by
\begin{eqnarray}\label{correl}
D(\alpha)=D_0 e^{-(\alpha-\alpha_0)/\alpha_1},
\end{eqnarray}
or $\alpha(D)=\alpha_0-\alpha_1\ln(D/D_0)$ in the case of a strong scatter, see lower insets in Fig. \ref{Fig5}, (a-f), with $D_0, \alpha_0,\alpha_1$ in Table I. In other words, $D$ and $\alpha$ are strongly (anti-)correlated. Interestingly, variations of the parameter $\alpha_1$ are rather small, $\alpha_1\approx 0.100-0.116$, while variations of $D_0$ and $\alpha_0$ are significant reflecting the model of correlations and $\lambda$. Values of $D$ are scattered over three orders of magnitude in the case of exponential and power-law correlations, and even four (!) orders, in the case of Gaussian correlations. It comes as a surprise that the largest scatter appears in the case of Gaussian correlations, where the random potential is smooth and shows the least local bias because of very short-range correlations, see panel (e) in Fig. \ref{Fig5}.

Furthermore, for a smaller scatter, $|\alpha-\alpha_0|\ll \alpha_1$, (\ref{correl}) yields 
\begin{eqnarray}\label{linear}
\alpha(D)=\alpha_2-b_1D
\end{eqnarray}
with $\alpha_2=\alpha_0+\alpha_1$, $b_1=\alpha_1/D_0$. Indeed, the upper insets in Fig. \ref{Fig5}, (a,c,e) display such a linear dependence for $10<t<1000$, however, with $\alpha_2,b_1$, which are not related to $\alpha_0,\alpha_1,D_0$, see in Table \ref{TableI}.

Next, the ensemble-averaged time-averaged (EATA) variance of particle positions lies essentially below
the corresponding ensemble-averaged variance, cf. Fig. \ref{Fig5} (a,c,e). Also the corresponding values
of $\alpha$ are different. For example, in the upper inset in (a), $\alpha_{\rm EATA}\approx 0.512$ for
EATA,  while $\alpha\approx 0.571$ for the ensemble-average therein. Likewise, in the lower inset (a), 
$\alpha_{\rm EATA}\approx 0.727$ for EATA and $\alpha\approx 0.608$ for the ensemble-average.
Interestingly, a very similar feature, i.e.,  a small scatter in viscoelastic FBM regime ($t<10$ ns) with
$\alpha_{\rm EATA} \approx 0.52-0.54$ and a strong scatter with  $\alpha_{\rm EATA} \approx 0.79-0.82$ 
was revealed in a DSPC lipid-cholosterol mixture in Ref. \cite{JeonPRL12}, see also Fig. 10 in review
\cite{Metzler16}.

\begin{table}[h]
	
	\caption{\label{TableI}Parameters of fit in Fig. \ref{Fig5}}
	\begin{indented}
		\item[]\begin{tabular}{@{}llllll}
			\br
			Panel & $\alpha_0$ & $\alpha_1$ & $D_0$ & $\alpha_2$ & $b_1$\\
			\mr
			a & 0.2875 & 0.1112 & 6.7742 & 0.7364 & 0.2230 \\ 
			b & 0.7269 & 0.1022 &0.2560 &   &  \\
			c & 0.3082 &0.1161 & 7.1809 & 0.7959 & 0.2086 \\
			d & 0.5833 & 0.1000 & 1.2907 & & \\
			e & 0.6410 & 0.1007 & 0.2546 & 0.8725 & 0.2854 \\
			f & 0.4717 & 0.1024 & 3.60887 & & \\
			\br
		\end{tabular}
	\end{indented}
\end{table}

\begin{figure}[h]
\vspace{0.8cm}
	\centering
	\includegraphics[height=5.4cm]{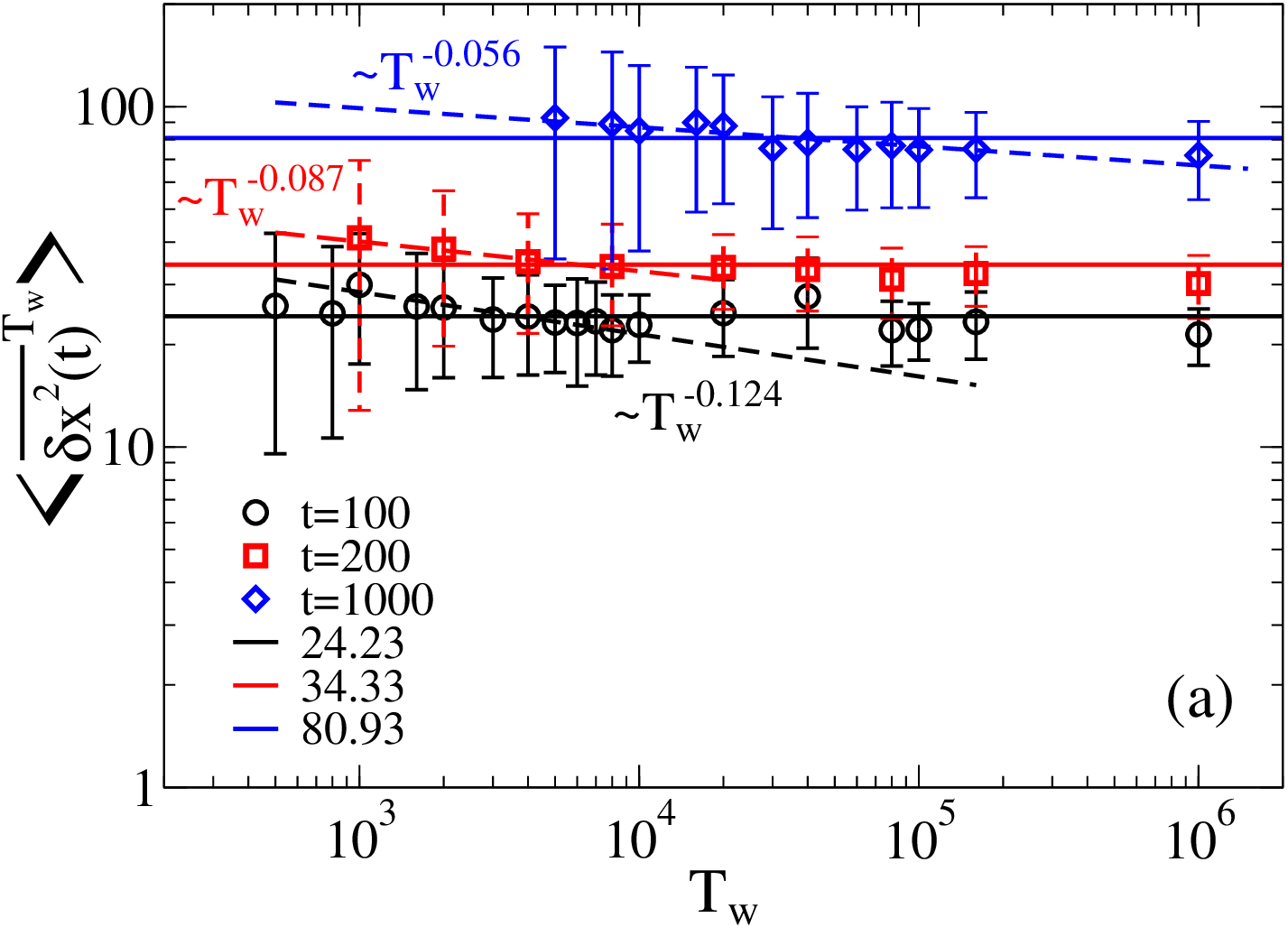} \hfill
	\includegraphics[height=5.4cm]{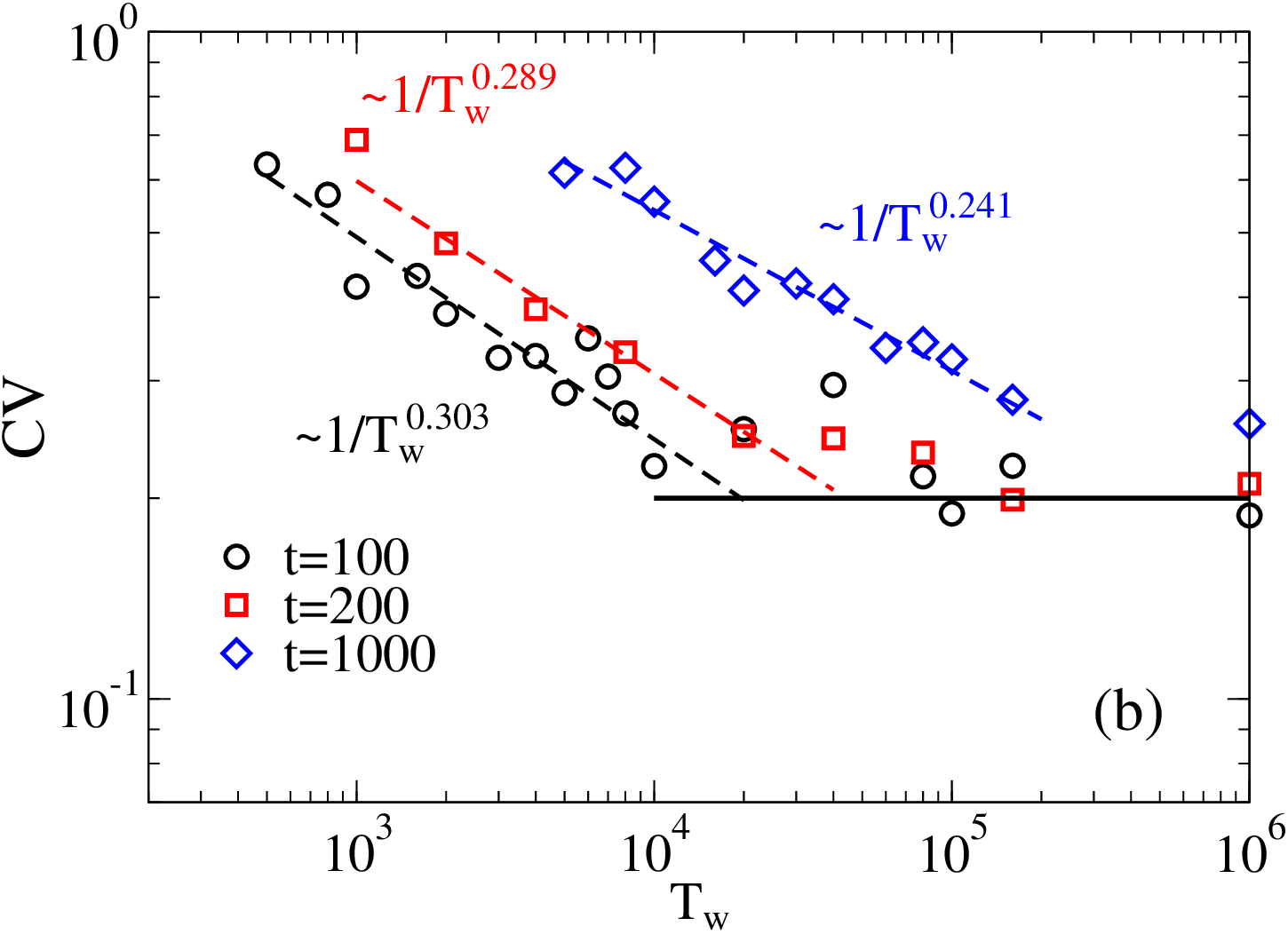} \vfill \vspace{0.5cm} \vspace{1cm}
	\includegraphics[height=5.4cm]{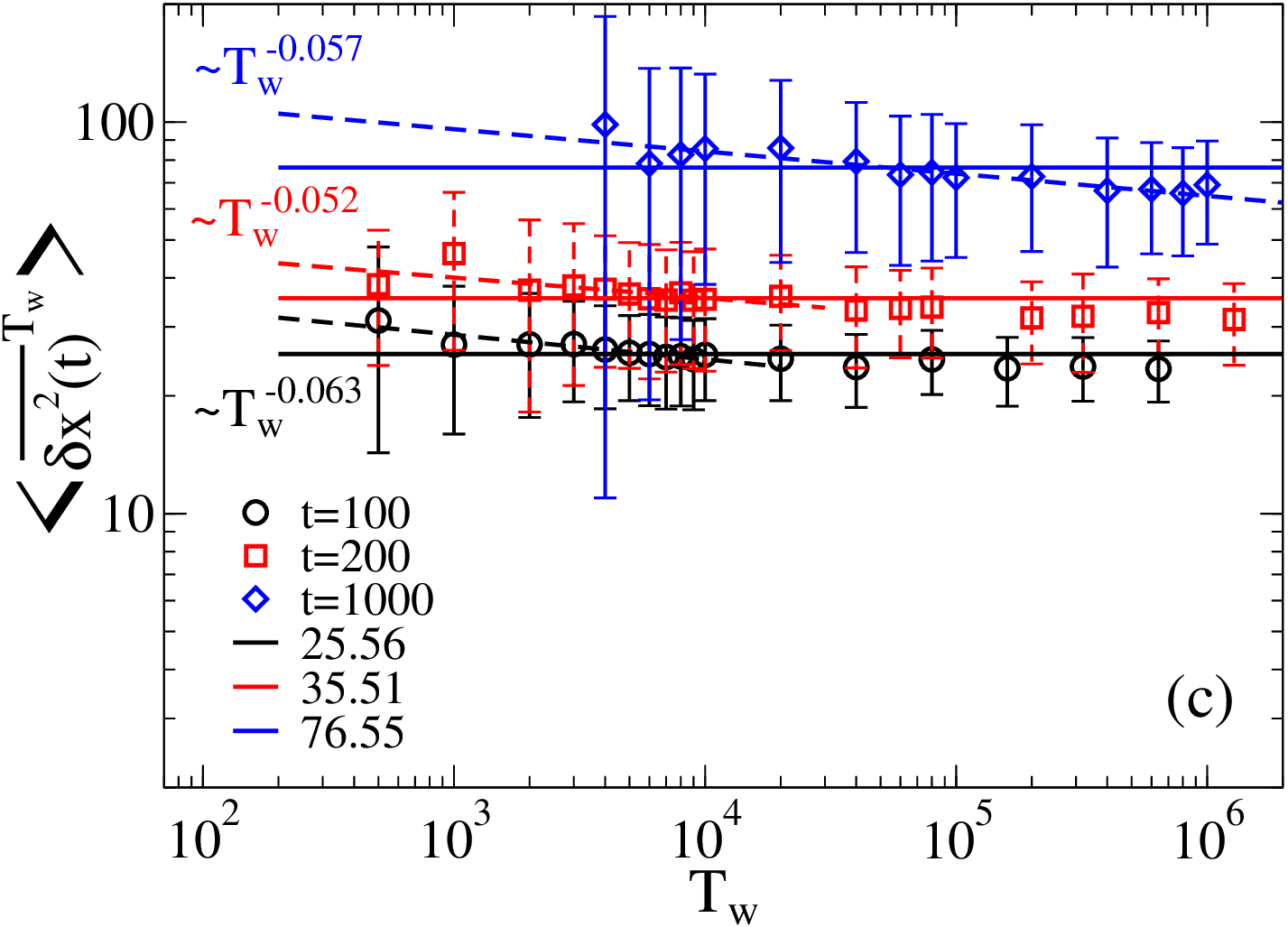} \hfill
	\includegraphics[height=5.4cm]{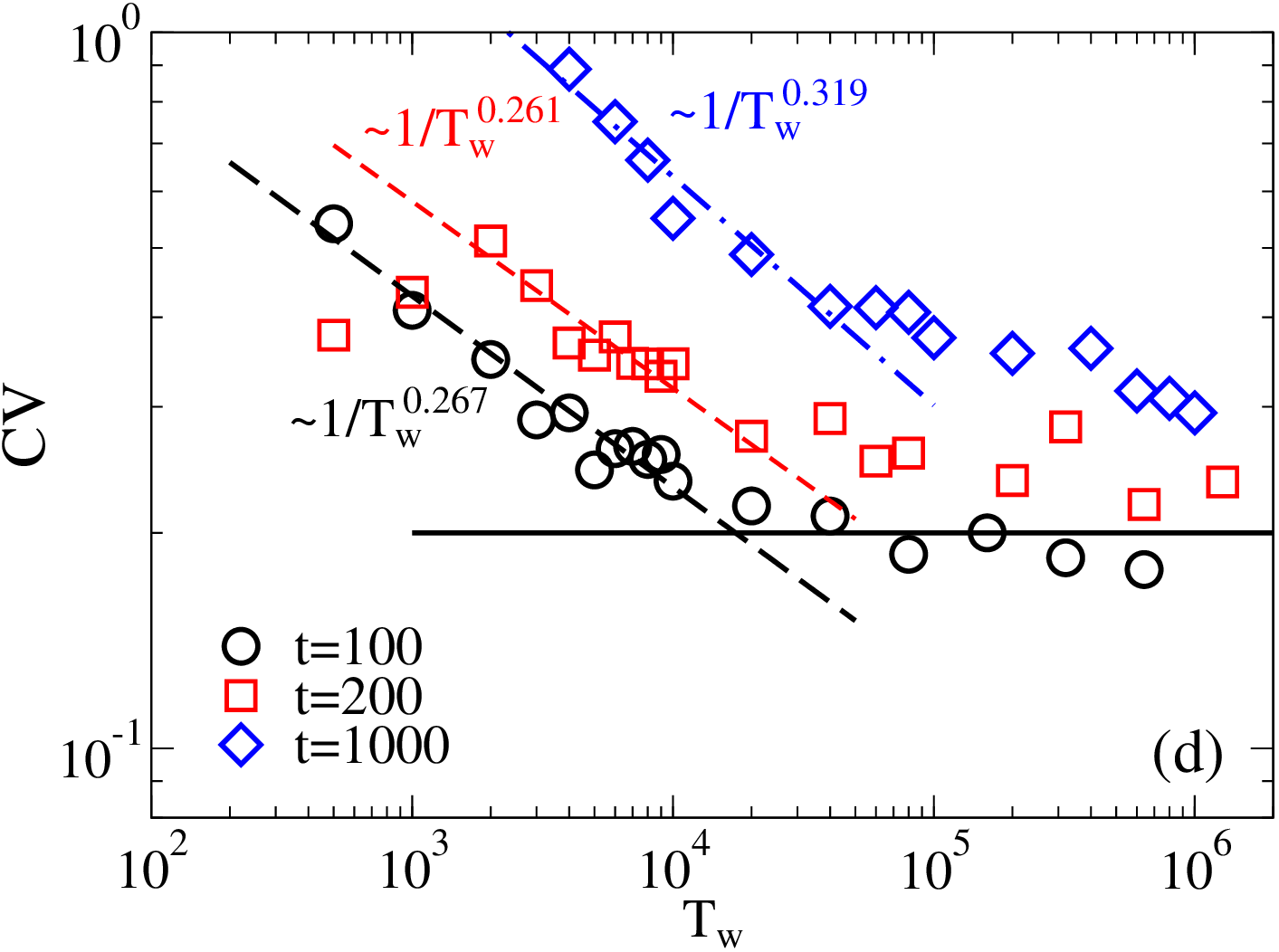} 
	\caption{ Dependence of EATA on averaging time-window ${\cal T}_w$  for three values of
$t=100,200,1000$ in the case of $\sigma=2$ for (a) exponential correlation  and (b) Gaussian
correlations. Panels (b) and (d) show the time-dependence of the corresponding coefficients of variation
(CVs) of the time-averages. The ensemble average over $M=200$ single-trajectory averages is done in each
case. Notice that a very weak ``aging'' in parts (a,c) occurs only until ${\cal T}_w \sim 100t$, see the
corresponding dashed lines $\sim 1/{\cal T}_w^a$, with power exponents $a$ shown in (a,c). Hence, this is
not genuine aging, but rather a transient statistics effect. The best fits with constants (full lines)
shown in (a,c) look almost perfect, upon taking error bars into account.  Also, CVs power-law decay with
exponent $a$ shown in (b,d) until ${\cal T}_w$ reaches about $100 t$ in each case. After this, CV
saturates around $0.2$ and remains significant. Hence, no self-averaging occurs with a further growing 
${\cal T}_w$. Time-averages remain random variables. Diffusion is not ergodic.	 }
\label{Fig6}       
\end{figure}

The case $\lambda=100$ implies a new feature. Here, practically no scatter appears until $t=10^3$ for
power-law and Gaussian correlations,  i.e., smooth disorder, cf. panels (d), and (f) in Fig. \ref{Fig5},
correspondingly. For rough exponentially correlated disorder, some very small scatter occurs yet for
$t<10^3$, see in part (b). However, in all cases, a profound scatter develops for $t>10^3$, see in
(b,d,f), especially insets therein. Here, the values $(D,\alpha)$ were extracted while applying the same
fitting procedure to a joint interval $[10, 2\times 10^5]$. Eq. (\ref{correl}) applies here with
parameters in Table \ref{TableI}.

\subsection{Aging}

Finally, we address the issue of ergodicity, its breaking, and the phenomenon of aging. The very fact
that the ensemble average and EATA are distinct implies that ergodicity is broken. Next,  the
ensemble-averaged $\left \langle \overline{\delta x_i^2(t)}^{{\cal T}_w} \right\rangle_M $ for a fixed
$t$ and sufficiently large $M=200$ shows first a power-law decrease with increasing ${\cal T}_w$ (a
spurious ``aging''), cf. Fig. \ref{Fig6}, a, c, for exponential and Gaussian correlations,
correspondingly. However, for a statistically significant ${\cal T}_w$, it just fluctuates around a mean
value (significantly different from $\langle \delta x^2(t) \rangle_M $) with root-mean-squared (RMS)
amplitude of fluctuations which very slowly decays with ${\cal T}_w$. The corresponding coefficient of
variation (ratio of this RMS to the mean value), square of which is named the ergodicity breaking
parameter or EBP within this context \cite{MetzlerPCCP}, $\rm EBP=CV^2$, is shown in Fig. \ref{Fig6}, b,
d. CV does initially decay as a power law, ${\rm CV}\sim 1/{\cal T}_w^a$ with $a\sim 0.241-0.319$.
However, this decay is only transient. For ${\cal T}_w>100 t$, it saturates around ${\rm CV}\sim 0.2$ or
${\rm EBP}\sim 0.04$. Diffusion is non-ergodic. However, aging is only transient and spurious. It can be
a typical experimental situation, upon a critical appraisal. 

Indeed, in Ref. \cite{TabeiPNAS} diffusion of insulin granules shows aging behavior in Fig. 2B therein.
However, maximal ${\cal T}_w/t$ in this figure is 100 (lowest curve therein). With a further increase of
${\cal T}_w$ (our notations are different), it saturates \cite{TabeiPNAS}. For the upper curve in the
discussed figure, the maximal value of ${\cal T}_w/t$ is 25.  Next, also a protein membrane diffusion in 
Ref. \cite{Manzo15} displays aging, at the first look, see Fig. 1 (f) in Ref. \cite{Manzo15} and also
Fig. 6 (f) in review \cite{Metzler16}. However,  a careful examination of the experimental data points in
these figures reveals that  $\langle D_{i, sgl}({\cal T}_w)  \rangle_M$ is nearly constant for ${\cal
T}_w\sim 0.33-3.33$ s evaluated for $M=600$ trajectories longer than 3.33 seconds (our notations are
different), contrary to the fit made, which overlooks this striking feature. It looks similar to one in
our Fig. \ref{Fig6} (a,c). 

\section{Conclusions}

The model of viscoelastic subdiffusion of finite temporal range in random Gaussian potentials developed
in this paper shows several remarkable features, which are consistent with many experimental observations
and with the viscoelastic nature of cytosol and lipid membranes. First,  it is consistent with fractional
Brownian motion anti-correlations revealed in many experiments. Second, experimentally claimed power-laws
are often even looking visually not quite as power laws (straight lines in double logarithmic
coordinates). A generalized log-normal distribution, which features systems with Gaussian disorder, can
provide a better fit in many situations \cite{PCCP18}. Third, Gaussian disorder yields a non-Gaussian
exponential power distribution of particle positions.  Similar distributions  are often measured
experimentally and might be confused for exponential Laplace distribution. Using a variable power
exponent $\chi$ instead of the  fixed $\chi=1$ value (Laplace distribution) in fitting experimental data
can help to reveal this remarkable feature, which can be often overlooked. 
Fourth, subdiffusion's viscoelastic nature is entirely consistent with a broad
scatter of single-trajectory averages, which can reach four orders of magnitude in anomalous diffusion
coefficient, even for a moderate disorder strength, $\sigma=2k_BT$. Simultaneously, the ensemble averages
are only weakly sensitive to the presence of disorder, which is also a vital feature. The fifth,
power-law exponent of subdiffusion $\alpha(t)$ can show non-monotonous time-dependence similar to one
found in molecular-dynamic simulations of crowed lipid-protein systems.  Sixth, single-trajectory
averages show a transient aging behavior and no aging when the time-averaging interval exceeds the
corresponding time more than a hundred times. However, scatter remains also for a further increasing
time-averaging interval. Hence, such diffusion is not ergodic. Ensemble and ensemble-averaged
time-averages are significantly different in general. However, aging is transient and spurious. In this
respect, typical published experimental data on aging is consistent with our observations, as we also
detailed in this work.  All these features are quite robust and universal concerning models of
correlation decay.

These profound features establish our model as a promising exploratory framework for subdiffusion in
complex fluid-like disordered environments. Of course, generalizations to other models of quenched
disorder and higher dimensions are highly desirable and welcome. 
We believe that the presented results, due to their generality,
will help to explain the bulk of pertinent experimental data and inspire a subsequent research work.

\section*{Acknowledgment} 

Funding of this research by the Deutsche Forschungsgemeinschaft (German Research Foundation), Grant
GO 2052/3-2 is gratefully acknowledged. We acknowledge also support by the Regional Computer Centre
Erlangen, Leibniz Supercomputing Centre of the Bavarian Academy of Sciences and Humanities, as well
as University of Potsdam (Germany), which kindly provided GPU high-performance computational
facilities for doing this work.

\appendix

\section{Numerical approach }\label{appendA}

We detail the numerical approach in this Appendix. It is based on approximating the power-law memory
kernel by a sum of exponentials (a Prony series expansion \cite{Prony,Hauer,Park99,Schapery99}) and
hyper-dimensional Markovian embedding of underlying non-Markovian dynamics \cite{GoychukPRE09,
GoychukACP12, SiegleEPL11,GoychukPRL19}. The method is numerically accurate.  The results coincide within
the numerical precision tolerance with the analytical results available in case of linear dynamics.

The memory kernel approximation reads \cite{GoychukPRE09,GoychukACP12}  
\begin{eqnarray}
\eta(t)= \sum^N_{i=1}k_i \exp\left(-\nu_i t\right),
\label{Prony}
\end{eqnarray}
where  $k_i= C_\alpha(b)\eta_\alpha\nu_i^\alpha/|\Gamma(1-\alpha)|$,  and $\nu_i=\nu_0/b^{i-1}$. The sum of exponentials obeys a fractal scaling with a scaling parameter $b$. It approximates the power-law decay \cite{Hughes, PalmerPRL85, GoychukPRE09, GoychukACP12} of this memory kernel in an almost optimal way \cite{Bohud07}. The choice of $\nu_0$ is related to the time step of simulation $\Delta t$. To avoid numerical instability, $\nu_0 \Delta t$ should be smaller than one. The power-law regime extends in this approximation from a short time (high-frequency) cutoff, $\nu_0^{-1}$, to a large time  (small frequency) cutoff, $\tau_h=\tau_l b^{N-1}$. The choice of $N$ is dictated by the maximal time $t_{\rm max}$ of simulations: $\tau_h$ should exceed $t_{\rm max}$ by at least several times, if one wishes to simulate FLE dynamics on the whole time scale. 

\begin{figure}[h]
\vspace{0.3cm}
	\centering
	\includegraphics[height=5.2cm]{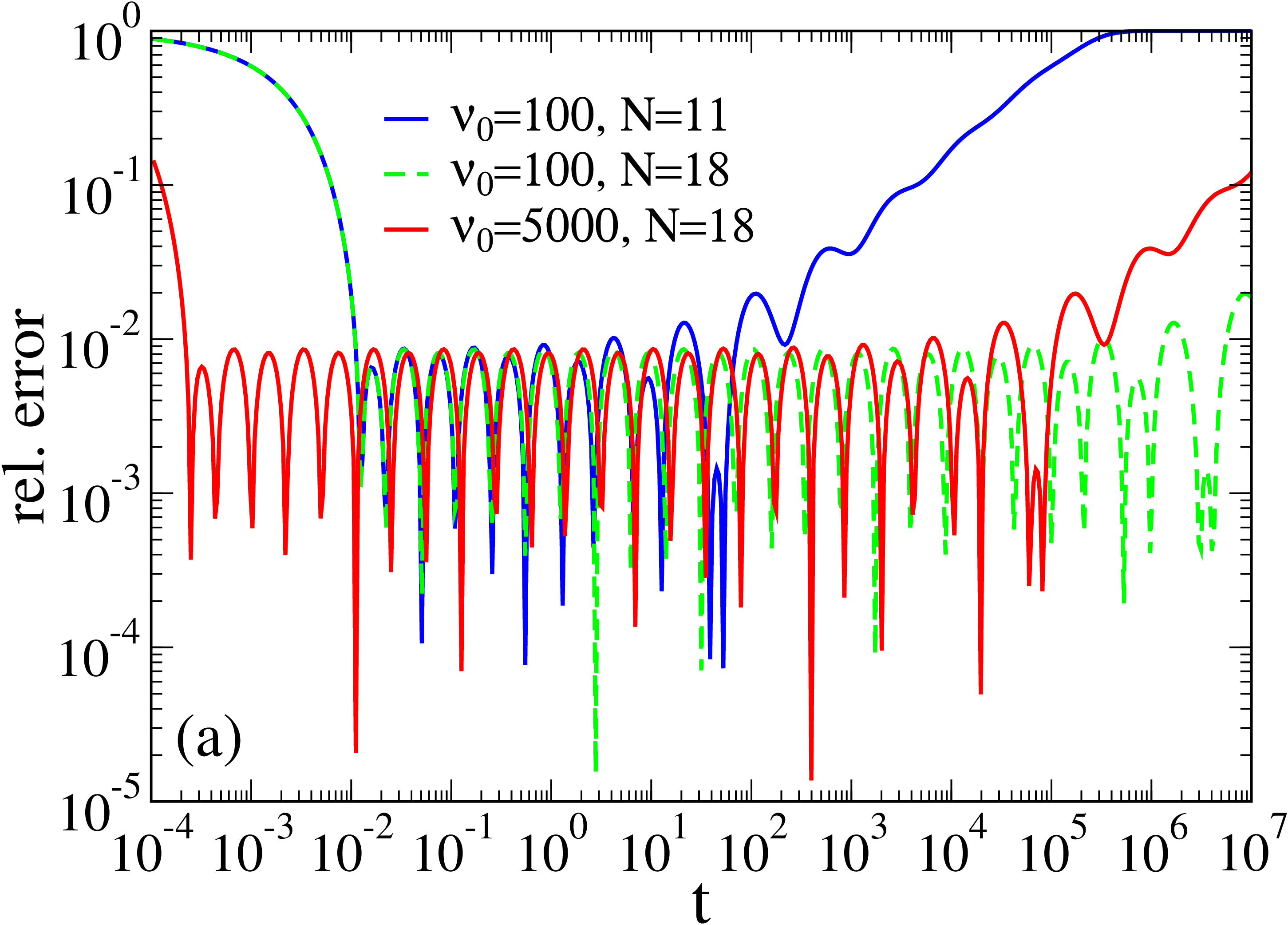} \hfill
	\includegraphics[height=5.2cm]{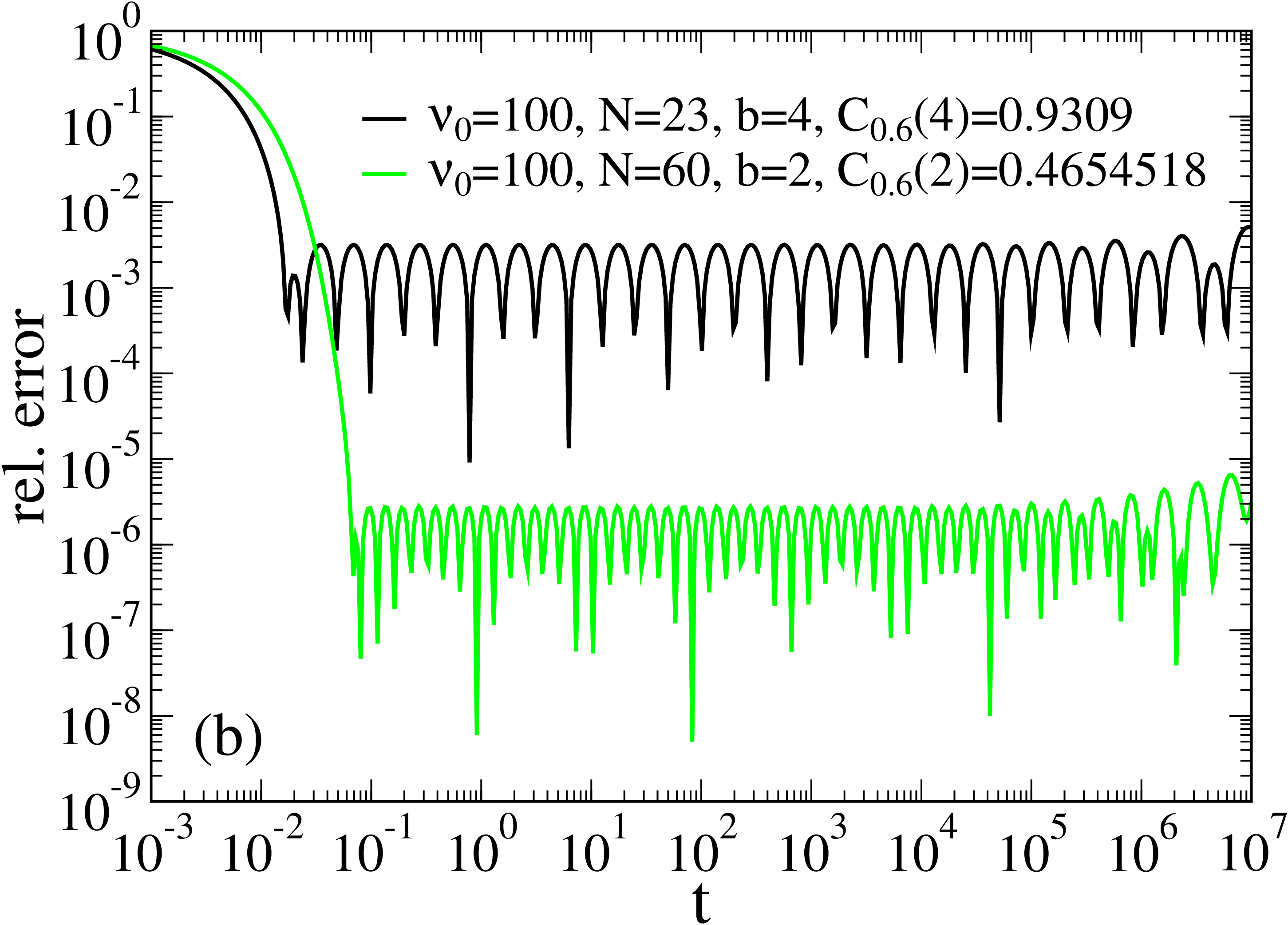}
	\caption{Relative error of the approximation (\ref{Prony}) to a strict power-law dependence $1/t^{0.6}$ for $b=5,4,2$ and several different $N$ and $\nu_0$ shown in the plot. In part (a), $b=5$, $N=11$ and $N=18$. Part (b) shows that with lowering $b$ to $b=2$ and increasing $N$ to $N=60$ one can increase relative accuracy of approximation to better than $10^{-5}$ or 0.001\%. It means that our embedding method can be made numerically exact in quite moderate embedding dimensions. Numerical accuracy of several per cents is, however, sufficient in most stochastic simulations.
	}
	\label{FigA1}       
\end{figure}

In this paper, we focus, however, on somewhat different setup. Namely, $\tau_h$
is quite finite and defines a maximal time range of GLE subdiffusion. In this
setup, subdiffusion occurring for $t\gg \tau_h$ is caused by the medium's
disorder. The accuracy of the approximation between two cutoffs is controlled by
the scaling parameter $b>1$. The smaller $b$, the better the accuracy. However,
a larger $N$ is then required. In this work, we use $\alpha=0.6$ in all
simulations, having in mind some lipid systems. With $b=5$ and
$C_\alpha(b)=1.0807$, approximation (\ref{Prony}) provides about 1\% accuracy
for $t$ between $10^{-2}$ and $10$ and $\nu_0=100$, see in Fig. \ref{FigA1}. It
still provides a sufficient accuracy of 5\% until $t=10^3$, i.e., over 5 time
decades. Hence, with time integration step $\Delta t=0.005$ in our stochastic
simulation it provides a good approximation of the corresponding FLE dynamics
until $t_c=10^3$. Then, the approximation becomes worse and for $t>t_c$ a
gradual transition to normal diffusion occurs. However, Fig. \ref{Fig2}, a and b
reveal that subdiffusion with $\alpha(t)\approx 0.6$ holds until $t=10^4$ in the
potential-free case. Increasing $N$ to $N=18$ provides a better than 1\%
accuracy from $t=10^{-2}$ until $t=10^6$. This choice has accuracy of 2\% until
$t=10^7$ -- the maximal time in our simulations. Moreover, choosing $b=2$ and
$C_{0.6}(2)=0.4654518$ would provide numerical accuracy of power law
approximation better than 0.001\% over more than eight time decades for a quite
moderate $N=60$. This is the reason why approximation in Eq. (\ref{Prony}) with
$\nu_i\propto 1/b^i$ is practically much better than a concurrent expansion
based on a scaling pertinent to eigen-mode expansions of polymer dynamics with
$k_i=const$ and $\nu_i\propto i^p$, with some constant $p$ \cite{McKinley}. For
example, in the case of Rouse model of polymeric chain, $p=2$. The latter one
needs, e.g., about $N=100$ of polymeric eigen-modes to achieve an acceptable
power law  approximation just over two time decades. Numerically, it is simply
not feasible \cite{PCCP18} to cover much larger time scales, like in  our
research, including this paper. 

\begin{figure}[h]
\vspace{0.4cm}
	\centering
	\includegraphics[height=5.0cm]{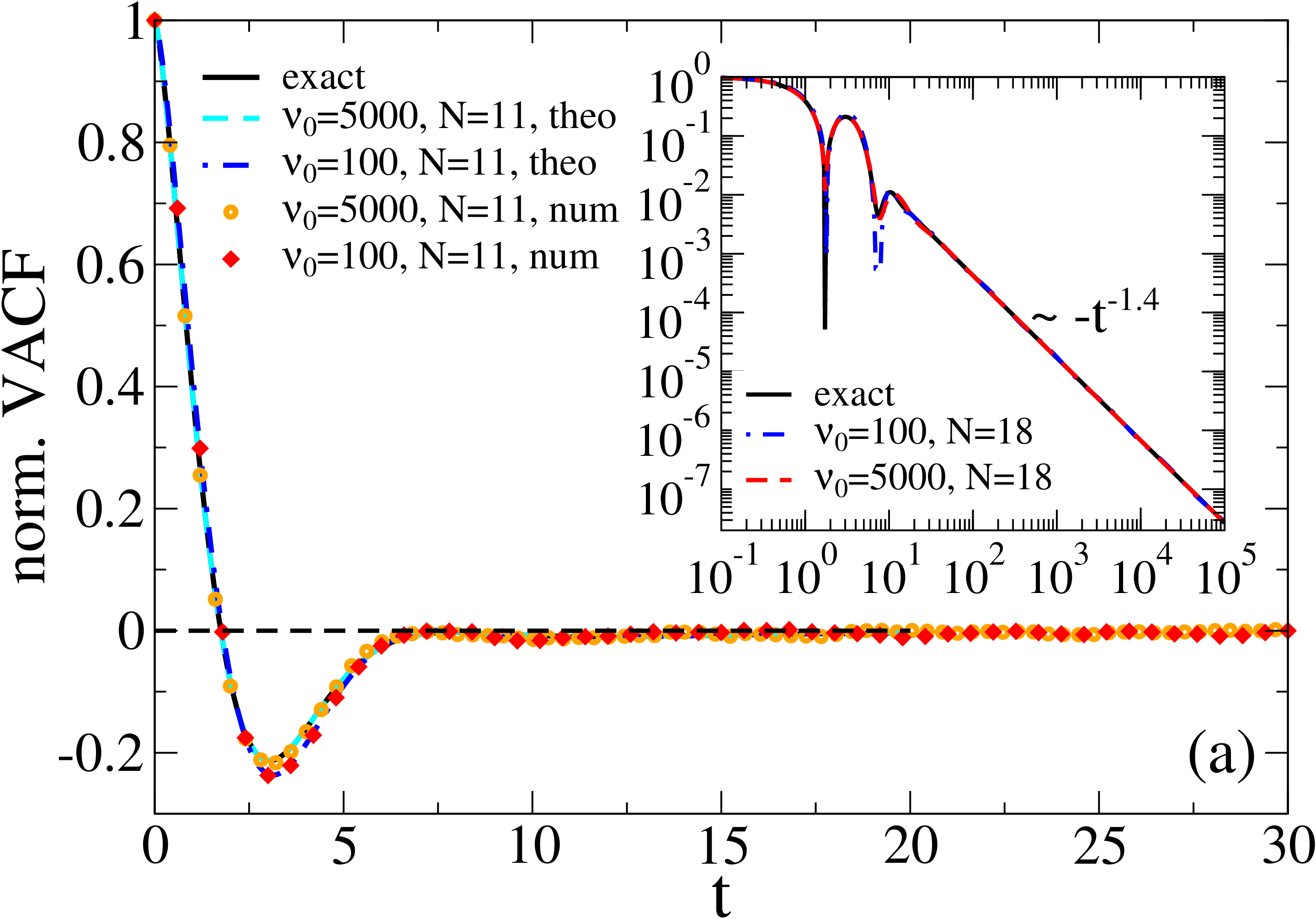}\hfill
		\includegraphics[height=5.0cm]{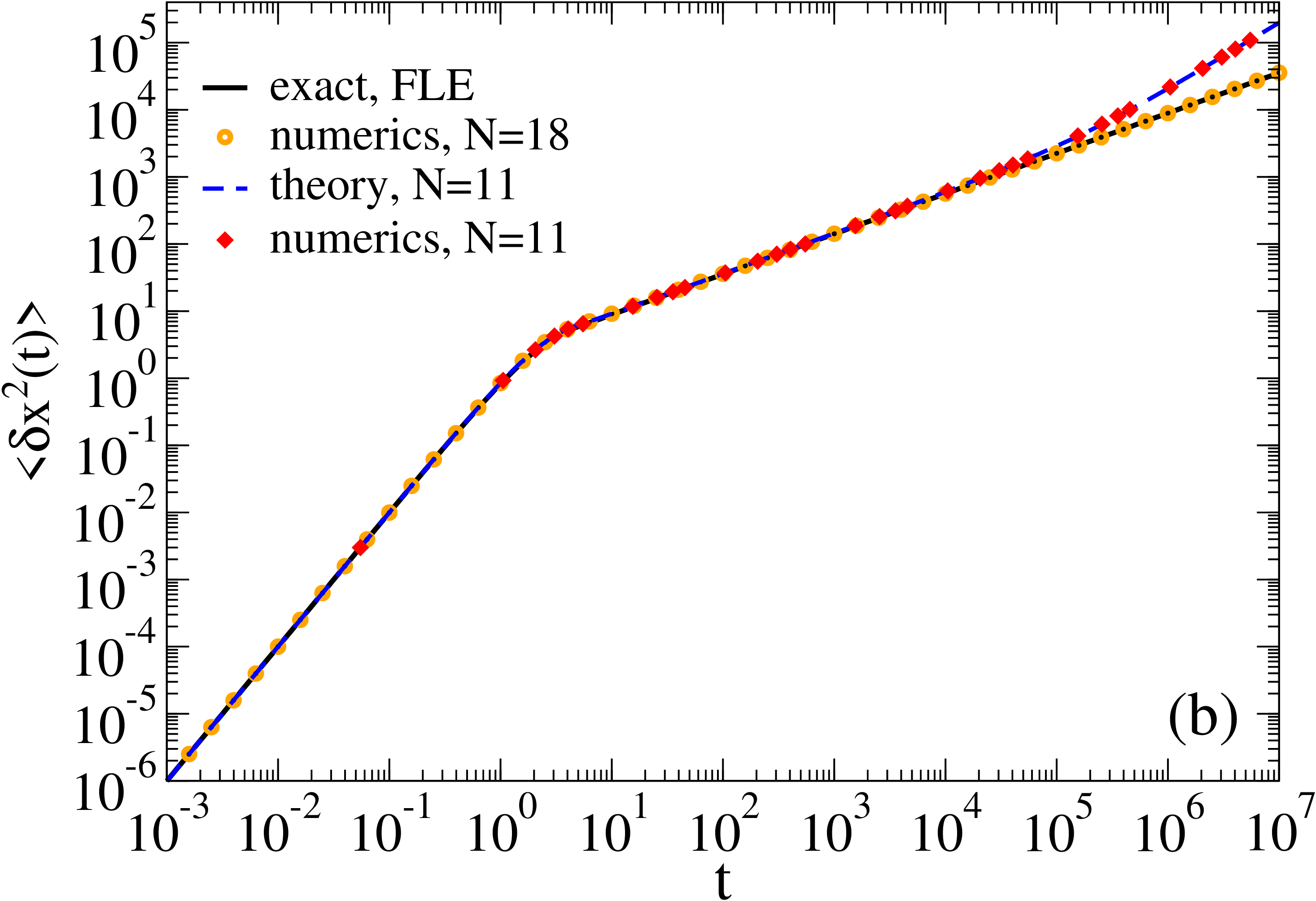}
	\caption{(a) Normalized velocity autocorrelation function (VACF) and (b) variance of the particle positions versus time for potential-free diffusion in the case $\alpha=0.6$. Exact analytical FLE results in Eqs. (\ref{Kv}) and (\ref{diff}) are compared with the results of various quasi-analytical approximations and matching numerical results. Quasi-analytical results are obtained by a numerically precise inversion of the corresponding Laplace-transformed analytical results containing approximate memory kernels (\ref{Prony}). Stochastic numerical results correspond to the same matching approximate kernels. They agree very well with quasi-analytical results for diffusion at all times, and also for VACF at sufficiently short times until statistical noise starts to dominate for small values of VACF. In the inset (a), the exact analytical result (\ref{Kv}) is compared with two quasi-analytical approximations. Here, the absolute value of VACF is plotted.
	}
	\label{FigA2}       
\end{figure}

For doing Markovian embedding, one introduces a set of $N$ viscoelastic forces $u_i$ such that 
the corresponding embedding dynamics in the hyperspace of dimension $D=N+2$ 
reads \cite{GoychukPRE09,GoychukACP12} 
\begin{eqnarray}\label{emb}
\dot x(t)&=&v(t)\nonumber \\
m\dot v(t)&=&f(x)+\sum^N_{i=1}u_i(t) \nonumber \\
\dot u_i(t)&=&-k_i v(t)-\nu_i u_i(t)+\sqrt{2k_BT k_i\nu_i}\zeta_i(t) ,
\end{eqnarray}
for $i=1,...,N$, where  $\zeta_i(t)$, $i=1..N$, are $N$ delta-correlated in time and mutually
uncorrelated white Gaussian noise sources of zero-mean and unit intensity, 
$\langle\zeta_i(t)\zeta_j(t^\prime)\rangle=\delta_{ij}\delta(t-t^\prime)$. The initial  $u_i(0)$ are
sampled as mutually independent Gaussian variables with zero mean and variances  $\langle
u_i^2(0)\rangle=k_BT k_i$ \cite{GoychukPRE09,GoychukACP12}. Eq. (\ref{emb}) corresponds to a
Maxwell-Langevin model of stochastic viscoelastic dynamics developed in our previous work
\cite{GoychukPRE09, GoychukACP12} and used in numerous papers. The first two equations in Eq. (\ref{emb})
are just equations of motion expressing Newton's second law. The last $N$ equations describe the
relaxation of viscoelastic force modes of environment influenced by the particle motion. Such an equation
corresponds to Maxwell's idea that viscoelasticity can be described within a framework of an elastic
force $u$ characterized by some elastic constant $k$, which can weaken in time and relax to zero with
rate $\nu$ \cite{Maxwell}. It is how Maxwell derived phenomenon viscosity from the phenomenon elasticity
in his classical work \cite{Maxwell} by course-graining over the times largely exceeding $1/\nu$.  The
last thermal noise term in Eq. (\ref{emb}) ensures that this viscoelastic force is consistent with
thermodynamics and FDT \cite{GoychukACP12}, i.e., it models a thermal environment at local temperature
$T$ and relaxes to equilibrium with rate $\nu$ for a localized particle, $v=0$. However, when a particle
moves, it disturbs this local equilibrium, what creates a long-lasting memory in the environment for
slowly relaxing modes $u_i$. These modes remember where the particle was before. Thereby, a slowly
relaxing local deformation is created, which attracts and temporary traps the particle (viscoelastic
caging). An analogy with polaron picture in solids is most useful here to grasp the physical mechanism of
viscoelastic subdiffusion \cite{GoychukPRE09, GoychukACP12}.  Notice that this approach is
straightforward to generalize \cite{SiegleEPL11,GoychukPRL19,GoychukPRE20} to include the effects of 
hydrodynamic memory \cite{FranoschNature,HuangNatPhys}, and a normal friction component \cite{GoychukACP12,PCCP18}.  The latter corresponds, e.g.,
the water content of cytosol. We do not consider them here. 

Initially, particles were always prepared with
their velocities Maxwell-distributed at temperature $T$ and localized sharply at some random, uniformly
sampled in the spacial interval $[-L/2, L/2]$ positions $x_i(0)$. It is a non-equilibrium initial
preparation. Particles are typically subjected to some local bias $f(x_i(0))$ and explore first some
local environment. They can, e.g., be either trapped in some local potential minima or start at the tops
of potential local hills, see in Fig. \ref{Fig1}.

In numerical simulations, we use the second-order stochastic Heun algorithm
\cite{GardBook} implemented in CUDA and run simulations on professional graphical processor
units (GPUs) with double precision. In ensemble simulations, we run $M=10^5$
particles in parallel. With $\nu_0=100$, $\Delta t=0.005$ and $N=11$
viscoelastic modes, simulations require about 6.5 days to arrive at $t_{\rm
max}=10^7$ on Titan V processors, and five times longer on older Kepler K20
GPUs. Embedding with $N=18$ on Titan V GPU requires about two weeks for the same
integration time step. It sets the practical restrictions in simulations.  We
first test our approach on calculating VACF $K_v(t)$ and $\langle \delta
x^2(t)\rangle $ in the potential-free case and comparing the results with exact
result in equations (\ref{Kv}), (\ref{diff})  for FLE dynamics. Results for
normalized $K_v(t)$ are shown in Fig. \ref{FigA2} (a). For $\alpha=0.6$,
$K_v(t)$ changes sign only once at $t_1\approx 1.76$ and remains negative for
$t>t_1$. It is different, e.g., from the case $\alpha=0.5$, where the sign
oscillations occur exactly three times \cite{PCCP18}. At $t_2\approx 3.04$ it
reaches first a minimum $K_v(t_1)\approx -0.21286\, v_T^2$, and then increases
and reaches  a local maximum  $K_v(t_1)\approx -0.00275\, v_T^2$ at $t_2\approx
7.88$, not crossing zero. Then, it again decreases to a local minimum
$K_v(t_1)\approx -0.00984\, v_T^2$ at $t_3\approx 11.88$, and after this it
gradually approaches zero in accordance with power-law $t^{-1.4}$. It has to be
mentioned that such fine features are not detectable in stochastic numerical
simulations using $M=10^5$ because of pure statistical relative error of the
order $1/\sqrt{M}\approx0.00316$ or 0.316\%. Because of this reason, the
numerical results in Fig. \ref{FigA2} (a) sink in the statistical errors already
for $t>6$. Hence, a trustful comparison of stochastic numerical and
(semi)-analytical results in part (a) can be made only for $t<6$. In part (b),
however, the agreement between the exact and (semi)-analytical results, from one
side, and numerical results, from another side, is nearly perfect for all times.
To resolve the discussed fine features in part (a), we also plot
(semi-)analytical results obtained by numerical inversion of the
Laplace-transformed 
\begin{eqnarray}\label{Laplace}
\tilde K_v(s)=\frac{k_BT}{ms+\tilde \eta(s)}
\end{eqnarray}
with the Laplace-transformed memory kernel approximated by Laplace-transformed
(\ref{Prony}). In this case, $\tilde K_v(s)$ is a rational function of $s$, and
it can be quasi-analytically inverted, while finding the roots of the
corresponding polynomial of $s$ in denominator of this rational function
numerically. However, for practical purposes it is best to use a Gaver-Stehfest
series \cite{Stehfest,StehfestErratum} for the inversion of Laplace-transform
with an adaptable (arbitrary in principle) numerical precision \cite{Valko}, as
we did earlier in many papers, e.g., in \cite{GoychukPRE09}. This inversion of
Laplace-transform allows us to arrive at numerically accurate results in many
cases, including exact and approximate memory kernel in this paper. In the
insert of Fig. \ref{FigA2} (a), where the exact (without cutoff)
$|K_v(t)/v_T^2|$   and its two approximations are plotted on a doubly
logarithmic scale, one can see that the approximation with $\nu_0=5000$ and
$N=18$ excellently reproduces $K_v(t)$ on the whole time span depicted. The
approximation with $\nu_0=100$ and $N=18$ also well reproduces all the features
mentioned above. However, it makes some discrepancies from the exact FLE result
near to the extrema of $K_v(t)$ in part (a) visible. Stochastic numerics in the
main plot in Fig. \ref{FigA2} for $\nu_0=100$, $\nu_0=5000$ (done in this case
with a much finer time-step $\Delta t=10^{-5}$), and $N=11$ also excellently
agree with (semi)-analytical results obtained by numerical inversion of
(\ref{Laplace}) [part (a)] and $2\tilde K_v(s)/s^2$ [part (b)]. The discussed
small discrepancies in a small intermediate range of $t$  in part (a) does not
influence the results already for  $t>12$, where the overdamped FBM regime is
established in the absence of external potential. For this reason, we used
embedding with $\nu_0=100$, which allows for a much larger $\Delta t$ and makes
feasible much large $t_{\rm max}$ in simulations. It is a typical trade-off
between numerical accuracy and feasibility in numerical simulations. Moreover,
we do not consider rigorously an FLE model in this paper, but rather a
subdiffusive GLE model with memory cutoffs.

\section{Estimation of physical parameters }\label{appendB}

In this estimation, we have in mind some lipid systems, where inertial effects
are clearly present in MD simulations on the initial time scale \cite{KnellerJCP11,JeonPRL12}. For example,
long-tail of VACF in MD simulations of a bilayer of 128
dioleoyl-sn-glycero-3-phosphocholine (DOPC) molecules at $T=310$ K, was fitted \cite{KnellerJCP11}
with $\alpha=0.61$, $v_T^2=6.55\times 10^{-3}\;{\rm nm^2/ps^2}$, $\bar
D_\alpha=0.101\;{\rm nm^2/ns^\alpha}$, yielding $v_T=0.0809\;{\rm nm/ps}$ and
$\bar\tau_v=(\bar D_\alpha/v_T^2)^{1/(1+\alpha)}=\tau_v/
[\Gamma(1+\alpha)]^{1/(2-\alpha)}=0.35$ ps or $\tau_v=0.323$ ps. Then,
$x_0=v_T\tau_v\approx 0.026$ nm or 0.26 \AA. In these units, $\lambda=10$ would
correspond to 2.6 \AA, and $\lambda=100$ to 2.6 nm. The maximal time $t_{\rm
max}=10^7$ in our simulations would correspond to 3.23 $ \mu s$. Notice,
however, that they are done not for a particular lipid or  lipid/protein system,
but present results of a generic physical model, and we are just playing with
parameters. They serve merely for readers' orientation. We use $\alpha=0.6$ in
simulations.

\vspace{1cm}

\bibliographystyle{iopart-num}

\bibliography{njp}

\end{document}